\documentclass[journal=jpcbfk,manuscript=article]{achemso} 

\usepackage{mathptmx}    
\usepackage{graphicx}
\usepackage{bm}
\usepackage{times}      
\usepackage{color}      
\usepackage{subfigure}  
\usepackage{amssymb}    
\usepackage{amsmath}    %
\usepackage{endfloat}    %

\title{Breakdown of inverse morphologies in charged diblock copolymers} 
\author{Monojoy Goswami}
\email{goswamim@ornl.gov}
\affiliation{Oak Ridge National Laboratory, Oak Ridge, TN, 37831} 
\author{Rajeev Kumar}
\affiliation{National Center for Computational Sciences, Oak Ridge National Laboratory, Oak Ridge, TN, 37831} 
\author{Bobby G.~Sumpter}
\affiliation{Oak Ridge National Laboratory, Oak Ridge, TN, 37831}
\author{Jimmy Mays}
\affiliation{Oak Ridge National Laboratory, Oak Ridge, TN, 37831}
\affiliation{Department of Chemistry, University of Tennessee, Knoxville, TN, 37996}

\begin{document}
\date{\today}
\begin{abstract}
Brownian Dynamics simulations are carried out to understand the effect of 
temperature and dielectric constant of the medium on microphase separation of charged-neutral 
diblock copolymer systems. For different dielectric media, we focus on the effect of 
temperature on the morphology and dynamics of model charged diblock copolymers. 
In this study we examine in detail a system of partially charged block copolymer 
consisting of 75\% neutral blocks and  25\% of charged blocks with 50\% degree of 
ionization. Our investigations show that due to the presence of strong electrostatic 
interactions between the charged block and counterions, the block copolymer 
morphologies are rather different than their neutral counterpart at low dielectric 
constant, however at high dielectric constant the neutral diblock behaviors 
are observed. This article highlights the effect of dielectric constant of two 
different media on different thermodynamic and dynamic quantities. At low 
dielectric, the morphologies are a direct outcome of the ion-counterion multiplet 
formation.  At high dielectric, these charged diblocks behavior resembles that of 
neutral and weakly charged polymers with sustainable long-range order.  
Similar behavior has been observed in chain swelling, albeit with small changes in 
swelling ratio for large change in polarity of the medium. The results of our simulations 
agree with recent experimental results and are consistent with recent theoretical predictions
of counterion adsorption on flexible polyelectrolytes.  
\end{abstract} 

\section{Introduction} 
\label{introduction} 
Microphase separation in neutral-neutral diblock copolymer melts and solutions 
has been studied extensively during the last two 
decades\cite{Bates1,Bates2,Russell1,Bates3,Fasolka1}. However, self-assembly of 
amphiphilic macromolecules still eludes a clear understanding. Despite the 
importance of these molecules for a number of technological and biological 
applications, a large parameter space affecting the self-assembly process poses 
serious problems for the scientific community.  

In particular, an amphiphilic diblock copolymer containing a charge block 
attached to a neutral block is of great importance to a large classes of nanotechnology 
applications ~\cite{Vriezema1,Zhang1,Jones1,Kim1,Ariga1,Li1,Eisenberg1} and
drug-delivery systems\cite{Gaucher1,Garrec1,Hu1,Savic1,Bajpai1,Beduneau1}.
Pioneer experiments on the self-assembly of amphiphilic copolymers in solutions by 
Eisenberg and his coworkers\cite{Eisenberg1} have revealed that different 
morphologies can be obtained by varying the composition of polystyrene 
(PS)-{\it b}-poly(acrylic acid) (PAA). In addition to the three well-known classical 
morphologies, namely, spheres, cylinders or rods and lamellae, two more morphologies 
were discovered. These two morphologies included vesicles in aqueous solution and 
simple reverse micelle-like aggregates in organic solvents. 

Origin of these morphologies (or micelles) in the context of amphiphilic copolymer 
solutions has been investigated extensively~\cite{Bates1,Bates2,Bates3,Jain1}. 
General consensus is that the micellization/aggregation of an amphiphile results 
from a balance between three different contributions to the free energy: chain 
stretching in the core, the interfacial energy and the repulsion/attraction between 
the coronal chains~\cite{Halperin2}. However, the presence of charges in the case of 
charged-neutral diblock copolymers complicates this simple picture of the balance of 
forces and sometimes, leads to non-trivial counter-intuitive morphologies.  For example, 
for a particular set of parameters, stacking disk-shape structures have been observed 
experimentally~\cite{Cui1,Li1,Jain1}.  As an another example, in a system consisting 
poly(acrylic acid-{\it b}-styrene) (PAA-{\it b}-PS) diblock copolymer, disk-shape 
one dimensional supra-assembly have been formed in a controlled manner~\cite{Cui1,Li1}. 

To understand the structures and the mechanisms by which amphiphilic diblock 
copolymers self-assemble into different morphologies, a comprehensive 
theoretical/simulation study is necessary.  However, theoretical work on the 
ordered morphologies of amphiphilic block copolymers is relatively 
scarce~\cite{Li3,Ortiz1,Guo2,Guo1}. The few theoretical studies of diblock 
amphiphiles and triblock copolymers that have been conducted to understand 
the formation of self-assembled structures includes molecular dynamics 
(MD)~\cite{Sknepnek1,Erdtman1,Srinivas1}, Monte Carlo 
(MC)~\cite{ChemPhysLett,Deschenes1,Du1} and self-consistent field theory
(SCFT)~\cite{Ma1,Sevink1}. Although most of these recent simulation studies
address self-assembly in solution~\cite{Guo2,Guo1}, recently there have been efforts
to approach the problem using dissipative particle dynamics (DPD) to understand
the mesoscopic self-assembly of amphiphilic block copolymers in polymer
melts~\cite{Li,Kriksin1,Khokhlov1,Xin1}. In solution, Guo 
{\it et.al.}~\cite{Guo2,Guo1} performed DPD simulation of paclitaxel loaded
poly(ethylene oxide)-{\it b}-poly(L-lactide) PEO-{\it b}-PLIA in water and
dimethylformamide (DMF) to observe self-assembled structures of
bicontinuous, lamella, rod and spherical micelles and provided a complete
phase behavior of the same system. Quite recently, it has been shown
by Kriksin {\it et.al.}~\cite{Kriksin1} that the local chemical structure of
monomeric units can influence the global self-assembled morphology of the
amphiphilic melts. It is interesting to note that so much effort has been
invested to understand block copolymer assemblies, however, the explicit
inclusion of charged blocks is by and large missing in most of the simulations
of amphiphilic block copolymers. So far, the self-assembly of charged-neutral 
diblock copolymers has been studied using analytical techniques such as the 
random phase approximation~\cite{marko1,marko2,kumar} (RPA), the SCFT~\cite{kumar}
and the MC~\cite{ChemPhysLett}. In MC studies~\cite{ChemPhysLett}, only the 
thermodynamics of the system have been investigated due to the equilibrium 
limitations of the Monte Carlo technique which may do justice to 
understand the morphologies but lacks a thorough investigation that 
include dynamics.  Only recently, experiments\cite{balsara} have been performed 
to understand the microphase separation in the charged-neutral diblock copolymer melts. 
Theoretically, it has been predicted that the entropic cost of confining the 
counterions to charged domains as a result of microphase separation stabilizes the 
disordered phase against ordered ones for a large parameter range. Furthermore, the 
morphology diagram is highly asymmetric and the regime of stability of the 
lamellar phase gets enhanced. We must point out here that in the field theoretical 
models~\cite{marko1,marko2,kumar}, possibility of counterion 
adsorption~\cite{manning,muthu_shulan,muthu_count,muthu_kumar} with the 
lowering of the temperature was not considered and hence, the models 
ignore the effect of ion pair formation and their implications on the 
structure.  This, in turn, limits the applicability of the field theoretical 
models without ion-pair formation to high temperature regime close to the 
disorder-order transition. 

In this article, we investigate the dynamics and self-assembly of charged 
diblock copolymers at nanoscale. The diblock copolymers consist of 
a charged block and an uncharged block. We are interested in the morphology and 
dynamics of charged block copolymers near the melt monomer density. 
Simulation of more realistic conditions requires an `explicit' presence 
of the charge-counterion interactions. In order to allow for an efficient 
simulation but account for explicit Coulomb interactions between the charges 
we employ a Brownian dynamics (BD) technique with Kremer-Grest bead spring 
model polymer to understand the structural as well as dynamical properties 
of the system and investigate the parameter space to identify the most 
important physics issues. The interplay between the entropy of the system and 
electrostatic energy between the charged monomers causes the formation 
of specific morphologies. We have observed that the diblocks of 50\% charge states 
on the charged block of an amphiphile forms structures that are quite different 
from their neutral counterpart. The chain swelling and other thermodynamic 
properties show the system does not follow the normal polyelectrolyte 
behaviors either. Similar results have been observed in recent experimental 
investigations~\cite{Cui1,Li1,SoftMatter} for sulfonated polystyrene (sPS) 
and fluorinated polyisoprene (fPI) diblock copolymers. These new classes of 
charged block copolymers show promising new directions to understand and 
thereby fabricate novel functional materials which  can be applicable to 
rather diverse applications from drug delivery to molecular electronics. 
The paper is organized as follows: Next section describes the simulation 
methodologies taking into account both Lennard-Jones (LJ) and Coulomb 
interactions. The relevancy of different parameters used in this simulation 
is explained in this section too.  The results are discussed in third section 
and compared with existing simulation and experimental results. In 
section\ref{conclusion}, we conclude with a short description of the major 
findings and their importance to novel material designing.  

\section{Simulation Method} 
\label{simulation} 
We used Brownian Dynamics (BD) simulations to examine the morphology and dynamics of 
charged block copolymer chains in a melt of density, $\rho = 0.7\big/\sigma^3$. The simulations 
are performed for two different system of sizes $V=16\times16\times16\sigma^3$ and 
$2V=20.16\times20.16\times20.16\sigma^3$. For both the systems chain length of $N=64$ is 
used. Each chain contains 75\% uncharged block and 25\% charged block. 
The charged block has 8 charges 
of charge $+q$ on it, giving a 50\% degree of ionization on the backbone. The charges 
interact via Coulomb forces with equal number of counterions randomly dispersed in the 
system.  The initial configuration of the model system is randomly generated at the 
melt density. All the monomers in the system have mass $m_i$ and Lennard-Jones 
diameter, $\sigma$.  Polymer chains are modeled following the Kremer-Grest bead spring
polymer model~\cite{kremergrest2} in which bonded beads are connected by 
finitely extensible nonlinear elastic (FENE) springs represented by, 
\begin{equation} 
U^{\rm FENE}_{ij} = -0.5kR_0^2\ln\left[1-\left(\frac{r_{ij}}{R_0}\right)^2\right] 
\end{equation}
where $R_0 = 1.5\sigma$ is a finite extensibility and the spring constant,
$k = 37.5\epsilon\big/\sigma^2$, $\sigma$ being the monomer diameter. The FENE
potential in combination with the (excluded volume) repulsive interaction creates
a potential well for the flexible bonds that maintain the topology of the
molecules. The energetic interaction between the  beads is modeled by a truncated and 
shifted Lennard-Jones potential. The repulsive part of the potential 
is given by, 
\begin{eqnarray} 
U^{\rm LJ}_{ij} = 4 \epsilon_R \left[\left(\frac{\sigma}{r_{ij}}\right)^{12} 
    - \left(\frac{\sigma}{r_{ij}}\right)^{6}+1\right], & r_{ij} \le 2^{1/6}\sigma \\ \nonumber  
  = 0,  & r_{ij} > 2^{1/6}\sigma 
\end{eqnarray} 
And the attractive part of the potential is given by, 
\begin{eqnarray} 
U^{\rm LJ}_{ij}  = 4 \epsilon_A \left[\left(\frac{\sigma}{r_{ij}}\right)^{12} 
    - \left(\frac{\sigma}{r_{ij}}\right)^{6}+1\right], & r_{ij} \le 2.5\sigma \\ \nonumber  
 = 0, & r_{ij} > 2.5\sigma 
\end{eqnarray} 

where and $r_{ij}$ is the distance between two monomers and $\epsilon_R$ and 
$\epsilon_A$ are the repulsive and attractive energy parameters respectively 
for two different interactions described below.
Each monomer of the system interacts via a short-range repulsive potential
whose interaction strength, $\epsilon \approx \epsilon_R=1.0$. The short-range 
repulsive LJ potential is shifted and truncated with a cut-off distance,
$r^{\rm R}_{ij}\le 2^{1/6}\sigma$. In addition to this repulsive
interactions, blocks A-A and B-B are attractive to each other with interaction 
strengths, $\epsilon_{AA}=2.0$ and $\epsilon_{BB}=4.0$ respectively. The above choice 
derives from the fact that the natural tendency of the different blocks of a BCP is to 
avoid each other because of the presence of dispersive intermolecular forces which often 
results in similar blocks more attractive to each other than the dissimilar monomers~\cite{Ruzette}. 
The cross interactions for A-B with strength, $\epsilon_{AB}$, are then obtained using Lorentz-Berthelot 
mixing rules. The repulsive cutoff is used in conjunction with the attractive cutoffs for the AA and BB 
interactions with a cut-off distance, $r^{\rm attractive}_{ij}\le 2.5\sigma$. 
As the focus of this paper is to observe the effect of electrostatics and entropy on the 
self-assembly and dynamics of the chains, we vary the temperature of the system along with 
the electrostatic parameters. 

The long-range interactions between the charges on the chain and the counterions are modeled 
using explicit Coulomb potential, 
\begin{equation} 
U^C_{ij} = \frac{q_i q_j}{Dr_{ij}} 
\end{equation} 
where $D$ is the dielectric constant of the medium. $D$ may not be constant throughout the 
system if dielectric mismatch is considered, however, for the purpose of this paper $D$ 
is assumed to be spatially constant. Long range Coulomb interactions are accounted
for through the Ewald sum \cite{Ewald}. 
Temperature is the energy scale parameter that is varied for different sets of simulations. 
We introduce a second energy scale parameter which is the ratio of Coulomb and Lennard-Jones interactions: 
$\xi_B=q^2\big/(D\sigma\epsilon_R)$. For real experimental systems, $\xi_B \sim$1-100: 
we use two values of $\xi_B=2$ and $10$. The parameter $\xi_B$ is 
proportional to the Bjerrum length, $l_B$ and is a constant.  

The dynamics of the monomers are governed by the classical Newton-Langevin equation,
\begin{equation} 
m_{\rm i} \frac{d\vec{v}_{\rm i}}{dt} = -\vec{\nabla}U_{\rm i} - 
\Gamma \frac{d\vec{r}_{\rm i}}{dt} + \vec{W}_{\rm i}(t)
\end{equation}
where $U_{\rm i}$ is the net potential energy experienced by particle
$i$ and $m_i$ is its mass. $\Gamma$ is the friction coefficient
between the chain monomer and background solvent. $\vec{W_i}(t)$
represents a Gaussian `white noise' with zero mean acting on each
particle~\cite{gunsteren1,gunsteren2}. The last two terms couple the system
to a heat-bath where the `friction term'  acts as a heat sink
and the `noise term' acts as a heat source. The first advantage of
this scheme is that the natural MD integration time-steps
are larger, thereby permitting simulation on longer time scales. A second
advantage comes from the fact that on this time-scale, only the mean effect
of the stochastic forces acting on the system needs to be considered, leading to
the first order temperature relaxation law which in turn reduces the need of
an external thermostat. The dimensionless units are defined as follows,
$t^{\star} = t\big/\sqrt{m_i\sigma^2/\epsilon_R}$, $\rho^{\star}=\rho \sigma^3$,
$T^{\star}=k_BT\big/\epsilon$, $U^{\star} = U\big/k_BT$ and $r^*=r\big/\sigma$.

\section{Results and Discussion} 
\label{results} 

In earlier studies~\cite{SoftMatter,ChemPhysLett}, we have demonstrated that 
controllable morphologies can be achieved by changing charge states of the 
charged diblock copolymer or by changing the dielectric constant of the 
solvent. In an attempt to establish a connection between thermodynamics to 
the dynamics of the system, we have performed a detailed molecular dynamics 
simulation of charged polymer and investigate the dynamics of the BCP 
and counterions for different thermodynamic parameters. To understand the 
system size effect we studied two systems, $V$ and $2V$ described earlier. 
We also have varied the Coulomb energy parameter, $\xi_B$.  The interaction 
parameter, $\xi_B$ is inversely proportional to the dielectric constant, 
therefore $\xi_B=10.0$ and $2.0$ are used to represent low and  high dielectric 
constant systems respectively. The system size effect is studied for $\xi_B=10$ 
cases only. 

\begin{figure}[tbp]
\pagebreak 
\centering
\subfigure[$T^*= 0.05$]{\includegraphics[height=6.2cm]{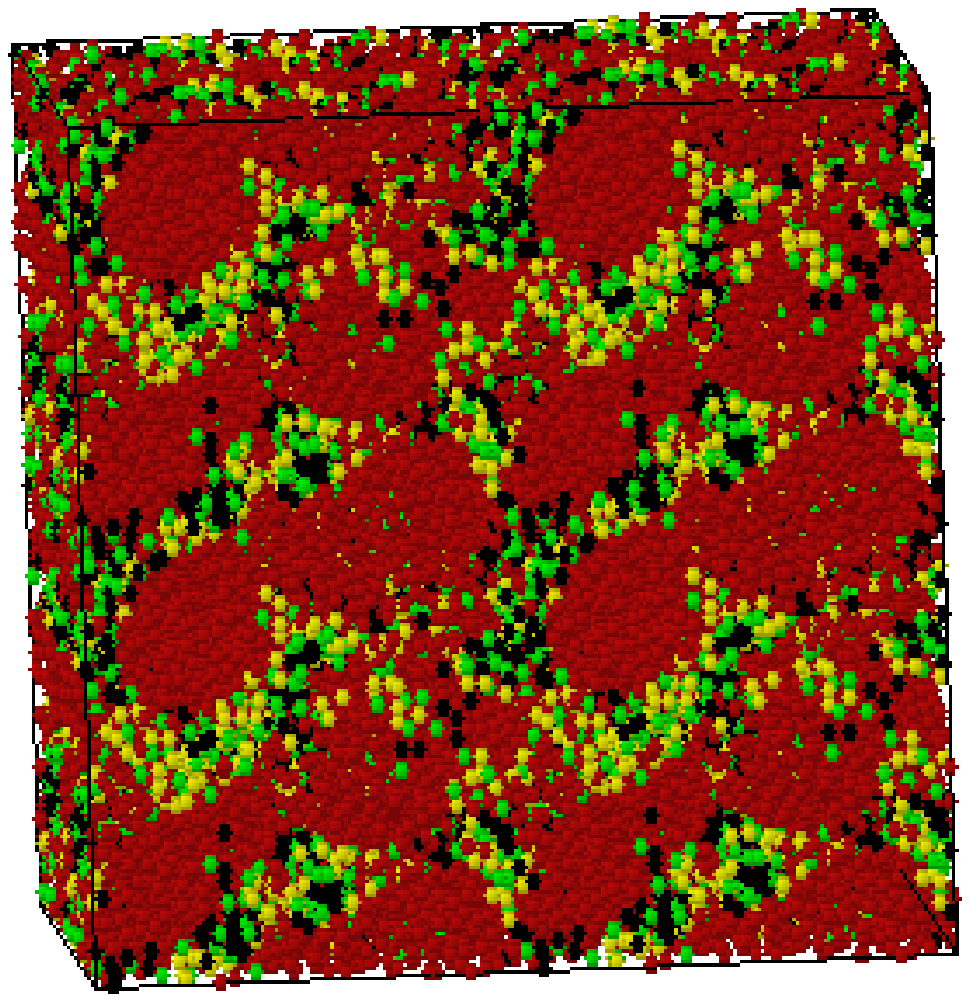}}
\subfigure[$T^*= 0.3$]{\includegraphics[height=6.2cm]{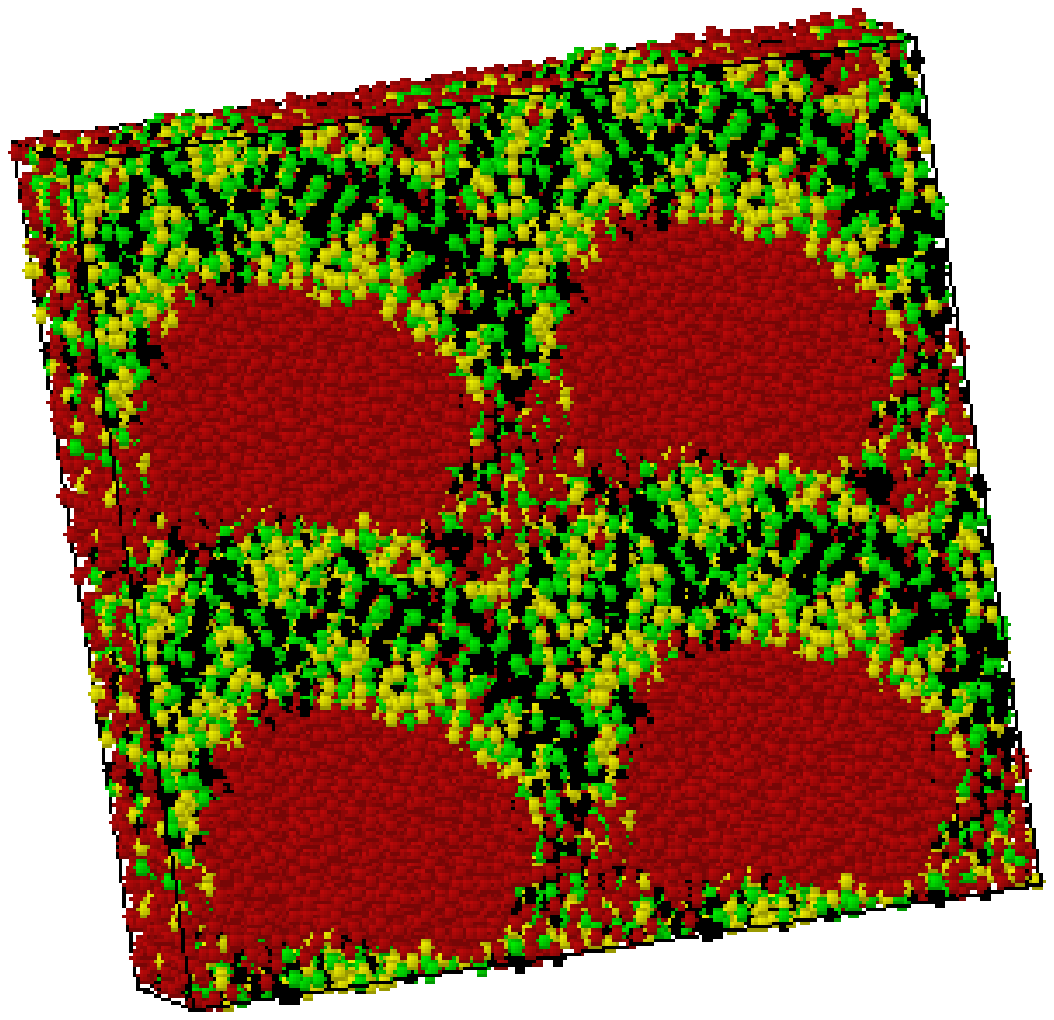}}
\subfigure[$T^*= 0.8$]{\includegraphics[height=6.2cm]{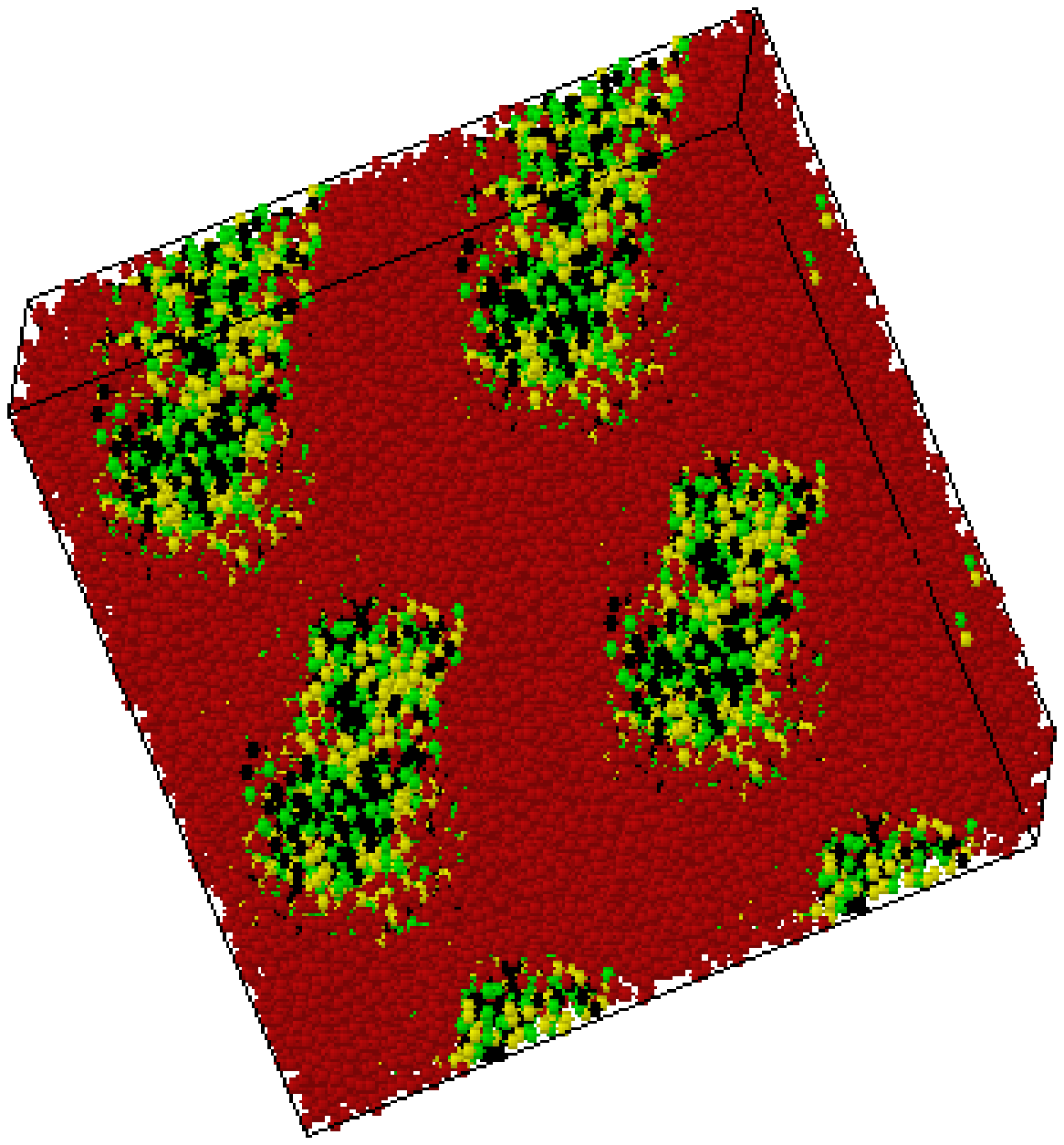}}
\subfigure[$T^*= 1.0$]{\includegraphics[height=6.2cm]{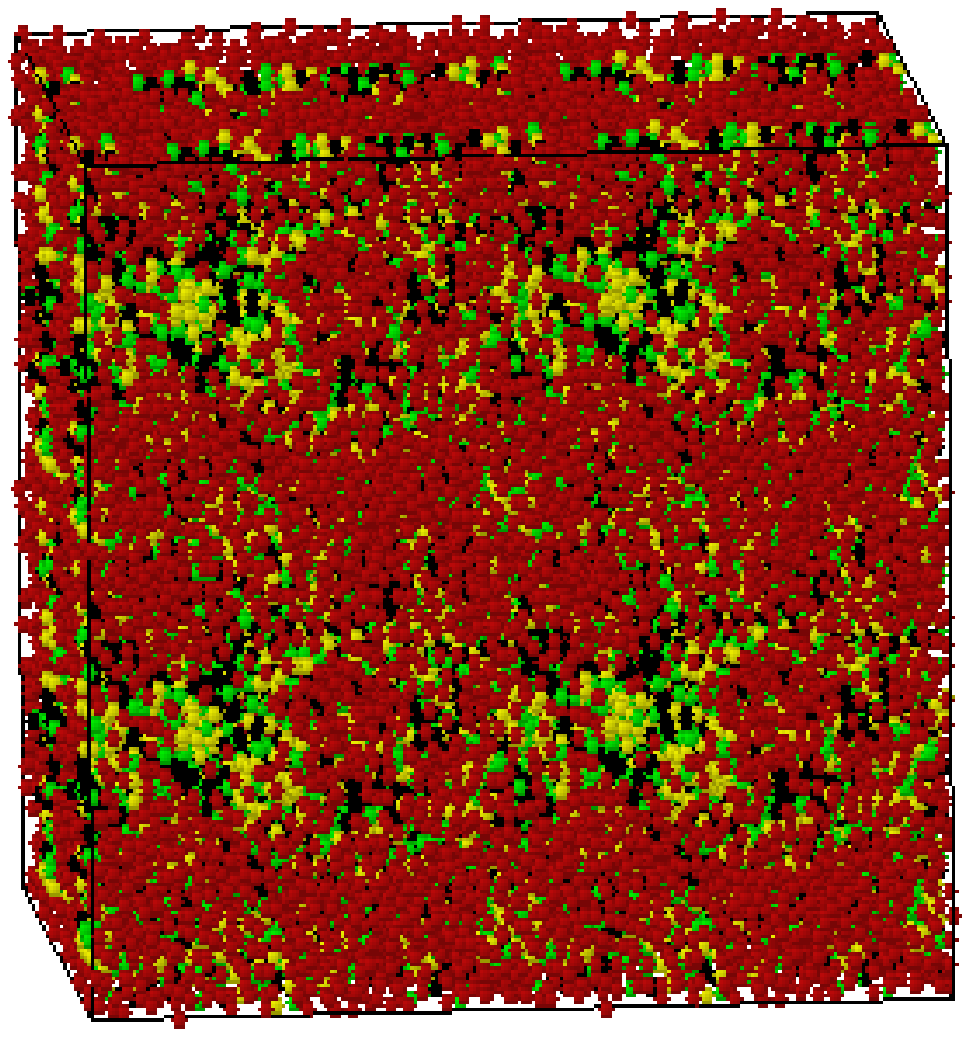}}
\caption{\label{snapshot} Snapshots at the end of the simulation 
for different temperatures at $\xi_B=10.0$ for $2V$ system. (a) $T^*=0.05$, 
(b) $T^*= 0.3$, (c) $T^*= 0.8$ and (d) $T^*= 1.0$. The maroon represents uncharged 
block monomers. Black and yellow represent charged block monomers, black being the 
neutral monomeric unit and yellow being the charged sites of the charged block. The green 
dots are the counterions. The central simulation cell is repeated in all the three 
directions to preserve the continuity of the chains images due to the periodic boundary 
conditions.}  
\end{figure} 

\ref{snapshot} shows the morphologies of the block copolymer melt 
for $2V$ system at $\xi_B = 10$. In these figures, the conventional 
diblock morphology as expected from the phase behavior of neutral block 
copolymers cannot be observed. From the morphology diagram for neutral 
diblock copolymers~\cite{Bates1,Bates2}, it is well established that the 
neutral counterpart of this system of 75-25 diblock would have exhibited 
hexagonally packed cylindrical morphology. In case of neutral diblock, the 
minority components form the hex structures and the majority blocks form the 
bulk. The presence of charge sites causes electrostatic interactions to dominate 
over the energetic repulsion effects between blocks that induces microphase 
separation in the neutral BCP. The strong electrostatic effect in the charged 
BCP changes the interaction energies such that the hexagonal structures cannot 
be observed. Lowering the temperature to $T^*=0.3$, gives rise to `inverse' 
morphology where the  minority blocks form the structures and the majority 
components form the matrix (\ref{snapshot}(b). At even lower temperature, 
$T^*=0.05$, the cylindrical nature of the `inverse' morphologies cannot be 
observed as can be seen in \ref{snapshot}(a). At the higher side of the 
temperatures, the charged blocks start to form well defined  structures 
(\ref{snapshot}(c)) that eventually settles down to entropy dominated 
miscible phase at very high temperature, $T^*=1.0$ (\ref{snapshot}(d)). 
The structures seen in \ref{snapshot} have also  been observed in recent 
experiments~\cite{SoftMatter}. The strong electrostatic interaction causes percolation 
of the charges that forms a network of charged blocks at low temperatures, giving 
rise to these unconventional `inverse' morphologies. Furthermore, note that the 
counterions are confined in the charged domains for all the temperatures corresponding 
to the ordered morphologies, in agreement with the field theoretical calculations on 
similar systems~\cite{kumar}. The lower the entropy of the system, stronger the effect 
of the electrostatics, giving freedom to the uncharged block to self-assemble to novel 
morphologies that would not have otherwise been possible.  

\begin{figure}[tbp]
\pagebreak 
\centering
\subfigure[$T^*= 0.05$]{\includegraphics[height=6.2cm]{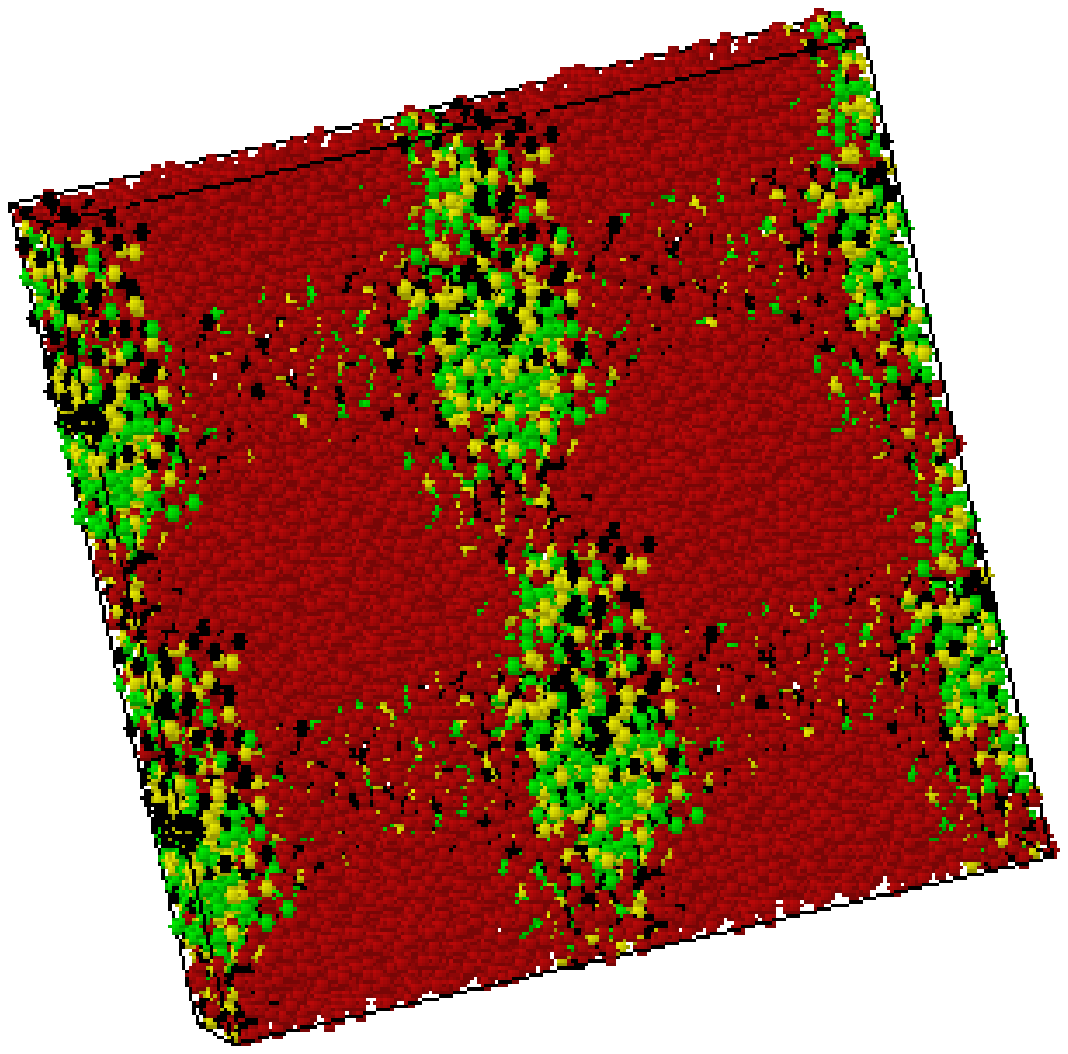}}
\subfigure[$T^*= 0.1$]{\includegraphics[height=6.2cm]{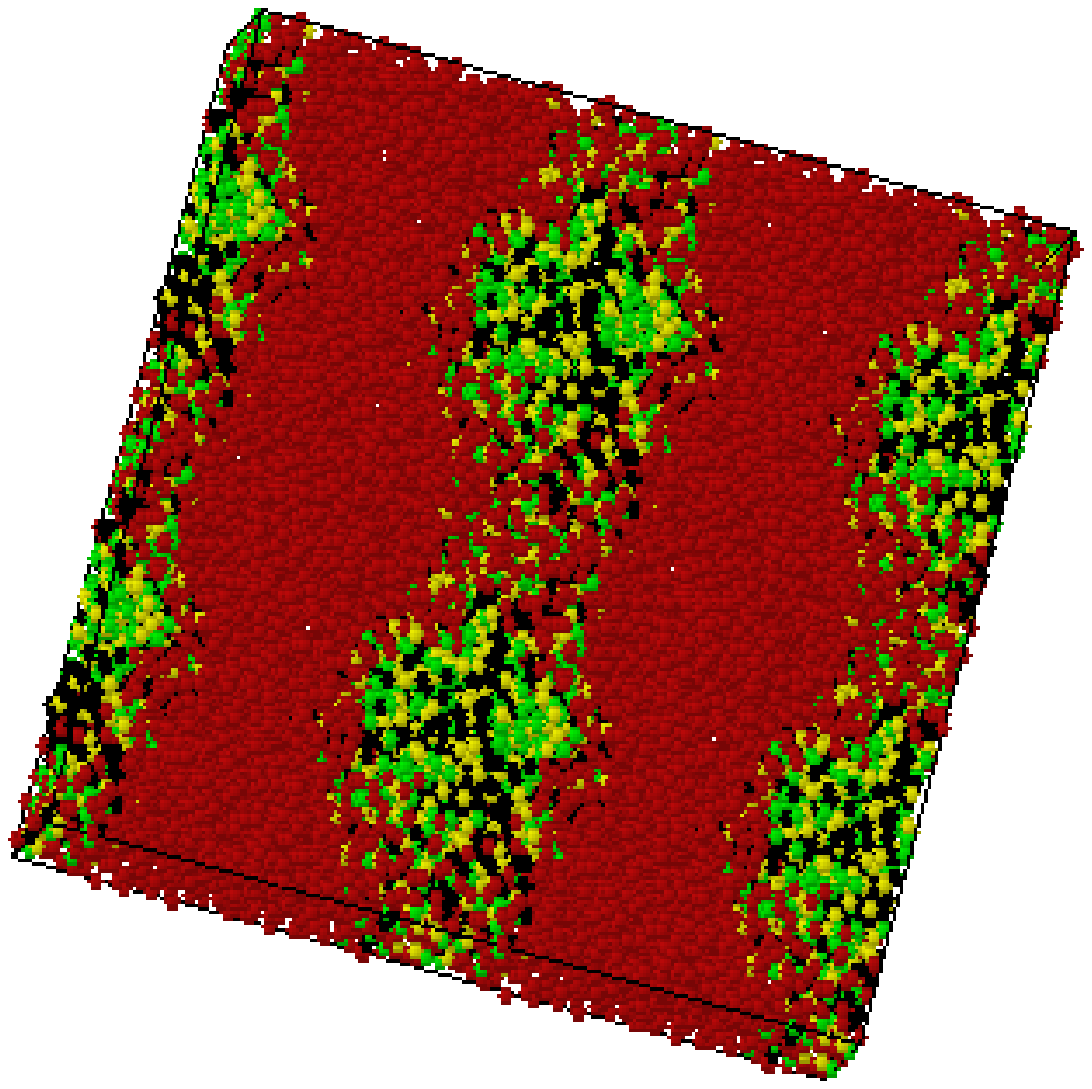}}
\subfigure[$T^*= 0.3$]{\includegraphics[height=6.2cm]{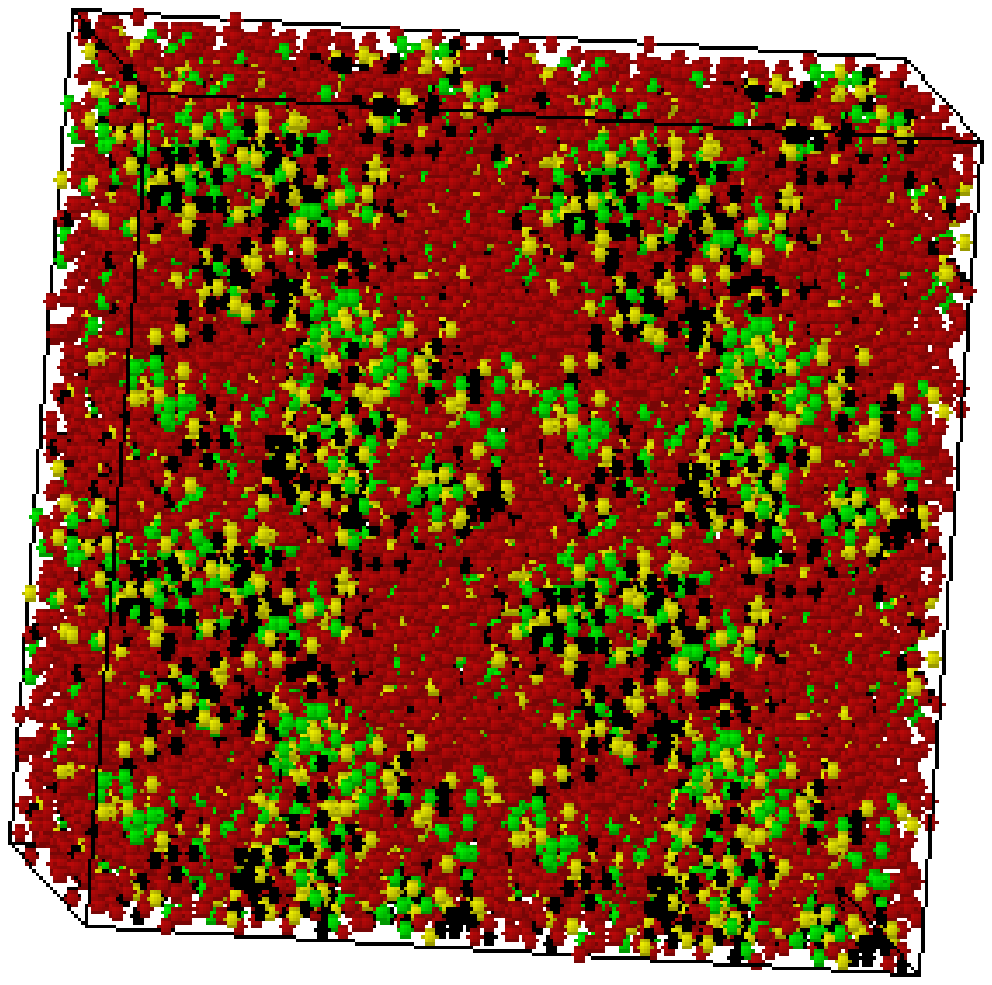}}
\subfigure[$T^*= 0.8$]{\includegraphics[height=6.2cm]{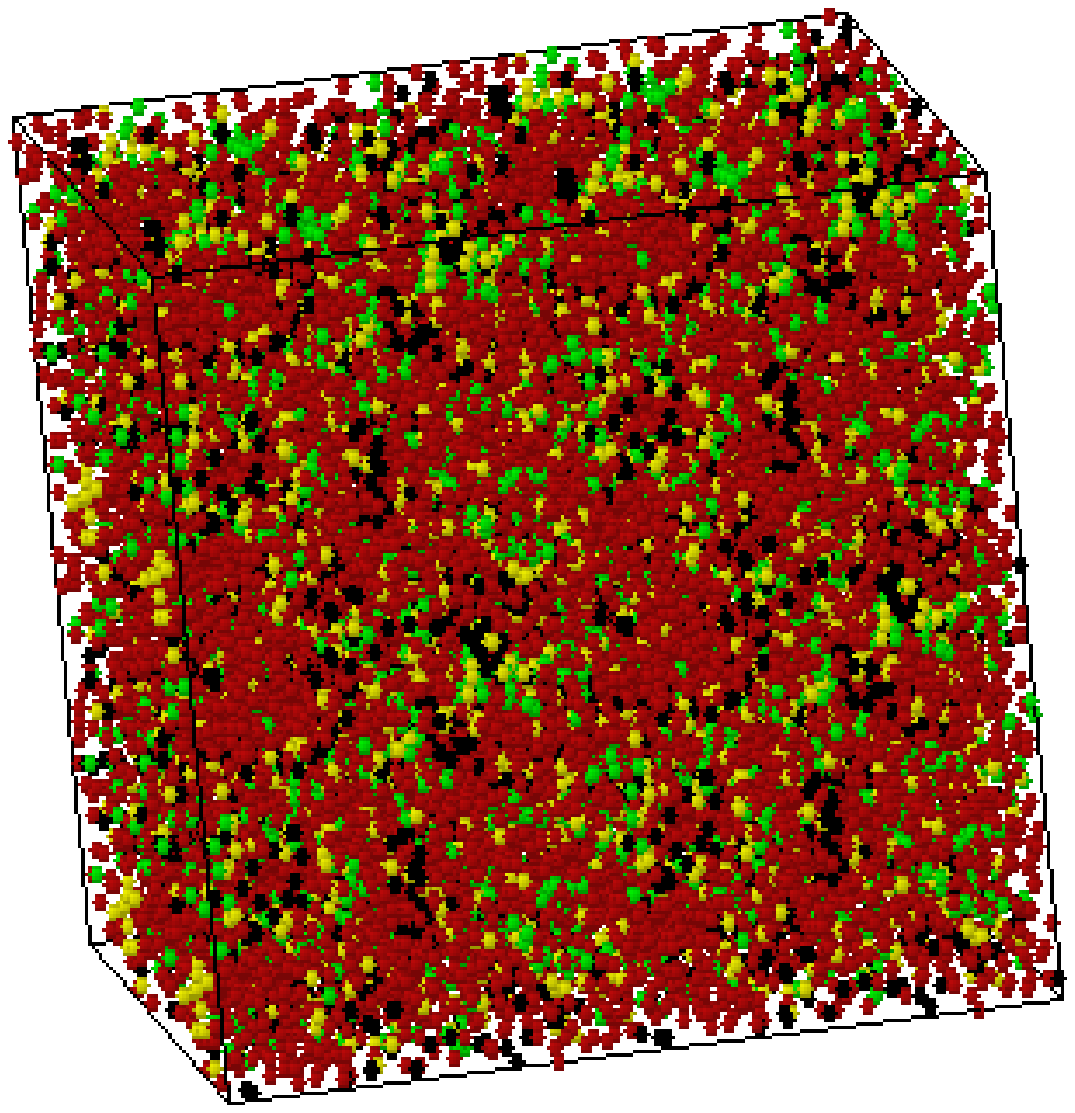}}
\caption{\label{snapshot_xi2} Snapshots at the end of the simulation 
for different temperatures at $\xi_B=2.0$. (a) $T^*=0.05$, 
(b) $T^*= 0.1$, (c) $T^*= 0.3$ and (d) $T^*= 0.8$. The maroon represents uncharged 
block monomers. Black and yellow represent charged block monomers, black being the 
neutral monomeric unit and yellow being the charged sites of the charged block. The green 
dots are the counterions. The central simulation cell is repeated in all the three 
directions to preserve the continuity of the chains images due to the periodic boundary 
conditions.}  
\end{figure} 

Snapshots for high dielectric constant at $\xi_B=2.0$ are shown in \ref{snapshot_xi2}.  
Interestingly these snapshots show similar behavior as neutral 75-25 diblock. 
At lower temperature end, in \ref{snapshot_xi2}(a) and (b),  the charged 
block (minority component) form the micro structures and the majority neutral blocks 
form the bulk. By increasing the temperatures the structures break down as can be 
seen in \ref{snapshot_xi2}(c) and (d) for $T^*=0.8$ and $T^*= 1.0$ respectively. 
At high dielectric constant (low $\xi_B$), the agglomeration of charges is more prevalent 
compared to low $D$ (high $\xi_B$). In a recent experiment~\cite{SoftMatter}, it has been 
shown that the long-range structures of charged diblock copolymers (sPS-fPI) breaks 
down by adding a trace amount of water in these system. It is well established for 
heterogeneous materials that an increase in dielectric constant results in a 
decrease in 
percolation and vice versa~\cite{Nan1}. It is interesting to observe the same physics 
dominating the microphase separation of macromolecules, in this case, the charged 
diblock copolymers. In \ref{snapshot_xi2}(a) and (b), increase in dielectric 
constant resulted in reduced percolation thereby allowing the charges 
(minority charged block) to agglomerate to form the structures whereas the 
majority block forms the bulk. 
A detail discussion on agglomeration of charges will be presented in 
connection with radial distribution function and fraction of free counterions 
later in this section. 
These results are in agreement with experiments~\cite{SoftMatter} and the SCFT 
calculations~\cite{kumar} carried out for weakly charged diblock copolymers. 
The addition of water increases the dielectric constant of the system which in turn 
lowers the electrostatic interaction strengths by a factor of $1\big/D$. This makes 
the percolation (responsible for `inverse' morphology) of charges more unstable, 
resulting in charge agglomeration (responsible for neutral BCP morphology) instead. 
Decreasing the electrostatic strength brings the system from an electrostatic 
energy dominated regime to an entropy dominated regime, where ion-pair formation is weaker 
compared to a percolated state. In the entropy dominated regime, the microphase separation 
is hindered by the entropic cost of partitioning of counterions in the charged domains 
as predicted by the SCFT calculations~\cite{kumar}. This also explains the breakdown/disappearance 
of `inverse' morphologies in \ref{snapshot_xi2} obtained at low $\xi_B$. In 
principle, one should obtain the `inverse' morphologies even for high $D$ ($\xi_B = 2$) 
medium by lowering the temperature further. However, it is extremely time consuming and hard 
to equilibrate the system at such low temperatures (with explicit Coulomb interactions).  

\begin{figure}[tbp]
\pagebreak 
\centering
\subfigure[$\xi_B=10$]{\includegraphics[height=6.2cm]{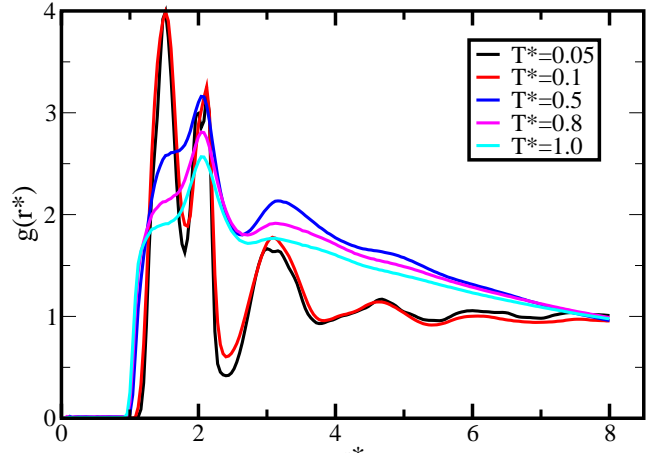}} 
\subfigure[$\xi_B=2$]{\includegraphics[height=6.2cm]{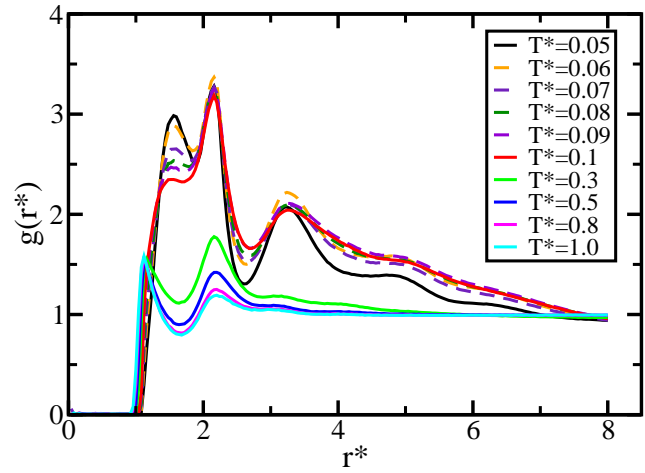}}
\caption{\label{rdfcion} Radial distribution function $g(r)$ of counterions 
for all the temperatures. The two different systems shown here are: 
(a) $\xi_B=10$ at $V$ system size and (b) $\xi_B=2$ for $V$ system 
size. For $\xi_B=2$, more temperature data have been obtained as shown in the 
legend} 
\end{figure} 

In order to corroborate the idea of ion-pair formation and its implications on the 
morphologies, we show the counterion-counterion radial distribution function 
(RDF), $g(r)$, in \ref{rdfcion}. 
For a complete understanding of ion-pair formation, a discussion of 
the ion-counterion RDF is extremely important. Due to the lack of data
of fixed charges (co-ion), the ion-counterion RDF cannot be presented
in this paper. Recent work on charged polymers~\cite{Goswami1} shows 
that the counterions always pair with ions. Ion-counterion pairing 
is also necessary from the electrostatics point of view, although this pair 
formation does not guarantee local electroneutrality. 
In \ref{rdfcion}(a) and (b) the RDF 
for $\xi_B=10$ and $\xi_B=2$ are shown respectively for the system size, $V$. 
We confirmed by plotting the RDF for $2V$ system that there are no apparent system 
size effects which might cause substantial error in the results presented in this paper. 
The system size effects for the simulation model can be neglected with confidence. 
One of the prominent features of these radial distribution functions can be seen by 
comparing the RDF for different temperatures. The first peak height increases by 
decreasing the temperature which represents increased agglomeration of counterions. 
The RDF shows a distinguishable decrease in agglomeration peak for $\xi_B=2$ compared 
to $\xi_B = 10$. This represents an enhanced miscibility shown in \ref{snapshot_xi2}. 
In \ref{rdfcion}(a) for low $D$, structures can be observed at temperatures 
as high as $T^*=1.0$ which is prominent in the snapshots shown in \ref{snapshot}. 
For $\xi_B=2$ system, an order-to-disorder like transition can clearly be observed from 
the sudden decrease in the $g(r)$ peak in \ref{rdfcion}(b). The transition temperature 
is hard to locate exactly, these plots can give a qualitative understanding of transition 
that is associated with charged diblock copolymers. In polymer melt, it had been 
shown~\cite{Lodge3} that the precise transition temperature was very hard to locate. 

For the model of a continuously and uniformly charged line, the radial distribution 
function implicitly has two peaks, the first peak corresponds to the 
condensed counterions and a second peak corresponds to the Debye-Huckel peak 
for the uncondensed counterions. The first peak is obviously an idealization of 
the real structure. In \ref{rdfcion}(a) and (b), the appearance of two peaks 
in RDF at low temperatures represents the counterion adsorption behavior akin 
to uniformly charged polyelectrolytes. However, this is not quite true because 
of the presence of a longer uncharged block, electrostatics and entropy both play 
a major role in deciding the morphology of the charged BCP. The increase 
in $T^*$ breaks the morphologies at high $D$ (\ref{snapshot_xi2}) which 
can also be seen in \ref{rdfcion}(b) where the first peak height is reduced. 
In \ref{rdfcion}(a) the peak heights are higher and represent stronger agglomeration 
compared to the high dielectric constant case, \ref{rdfcion}(b) which has 
lower peak heights representing a much weaker agglomeration, in this case the entropic 
effect is much more pronounced in forming the morphologies. In \ref{rdfcion}(a), 
the second peak becomes dominant which may result in longer inter-atomic separations in a 
low dielectric constant medium.  The first peak in \ref{rdfcion}(a) begins disappearing 
and the majority of counterions show agglomeration at an uncondensed phase (second peak) akin 
to the polyelectrolyte behavior.  Also at high $T^*$ the thicker width of $g(r)$ is representative 
of loosely bound counterions similar to polyelectrolyte behavior~\cite{muthu_shulan}. In 
\ref{rdfcion}(b), the first peaks shift to shorter inter-atomic separations with the 
second peak position unchanged. For high dielectric medium, $\xi_B=2$, the first peak positions 
shift due to tighter packing. 
The close packing of counterions seen in ~\ref{rdfcion} can electrostatically be 
possible if the counterions form charge-counterion pairs. Therefore, the neutral 
diblock behavior of the morphologies as observed in large $D$ (\ref{snapshot_xi2}) 
are due to the the tightly packed charge-counterion pair at each charge site 
of the polymer that causes the charge sites to resemble as neutral sites. 
On the other hand, the presence of the second peak indicates counterion-counterion 
correlations that extends throughout the charged block of the chain  which 
effectively shows apparent neutralization of the charged block. This relates 
quite well with the snapshots shown in \ref{snapshot} and \ref{snapshot_xi2}. 
From these morphological analysis and the structural evolutions for 
different dielectric constants, it can easily be concluded that   
at low $D$ the charged block exhibit polyelectrolyte behavior and at high 
$D$ it behaves like a neutral diblock. It should be noted that both the 
$\xi_B$ studied here have explicit point charges on the minority block of 
the chain. 

\begin{figure}[tbp]
\pagebreak 
\centering
\includegraphics[height=6.2cm]{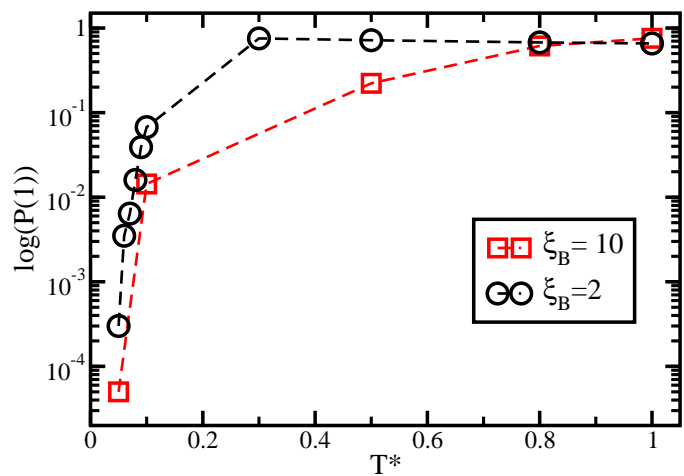}
\caption{\label{frac} Fraction of free counterions as a function of temperature for 
$\xi_B=2$ (black circles) and $\xi_B=10$ (red squares) systems. The lines in these 
plots are guide to the eyes.}  
\end{figure} 

To augment the understanding of charge agglomeration, cluster size distribution  
is investigated. A cluster is defined to be the total number of nonbonded charges 
associated with `a' charge within a cutoff radius $r_{\rm cut}$. The cut-off radius, 
$r_{\rm cut}$ is the location of the first minimum of RDF in the maximally clustered 
state. The total number, $N_c$, of non-interacting charge sites within $r_{\rm cut}$ 
is defined as the cluster size. The probability of occurrence, $P(N_c)$, of a 
cluster of size $N_c$ is computed. The probability of occurrence of cluster 
size $N_c = 1$, $P(1)$ is taken as the fraction of free counterions. In ~\ref{frac}, 
fraction of free counterions are plotted versus temperature for two different 
$\xi_B$. The red squares represnt $\xi_B=10$ and black circles represent $\xi_B=2$
respectively. Small $P(1)$ reflects lower free counterions, threfore higher 
agglomeration. For lower temperatures, very few counterions are free. Stronger  
agglomeration can be observed in snapshots at lower temperatures too. 
There is a sharp increase of $P(1)$ at around $T^*=0.1$ which may be associated 
with stronger agglomeration of charges. 
For $\xi_B=10$, the red sqaures show more clustering (less free counterions) 
compared to $\xi_B=2$. From the snapshots (\ref{snapshot} 
and \ref{snapshot_xi2}), it has been observed that 
low dielectric constant ($\xi_B=10$) shows more percolation of charges thereby 
forming reverse morphologies. These results are consistent with \ref{frac} in which 
it can be observed the $\xi_B=10$ forming more clusters (less free counterions). From 
these plots though the formation of charge multiplets cannot be guarateed as that 
requires an explicit representation of the fixed charge data which is absent 
due to unavailability of fixed charge postions. However, agglomeration of counterions 
cannot be justfied, from a electrostatics point of view, without the presence of 
opposite charges. For charged polymer systems, it has been observed 
earlier~\cite{Goswami1} that multiplets form in charge agglomeration. 

\begin{figure}[tbp]
\pagebreak 
\centering
\subfigure[Specific Heat, $C_V=(<E^2>-<E>^2)/T^{*2}$]{\includegraphics[height=6.2cm]{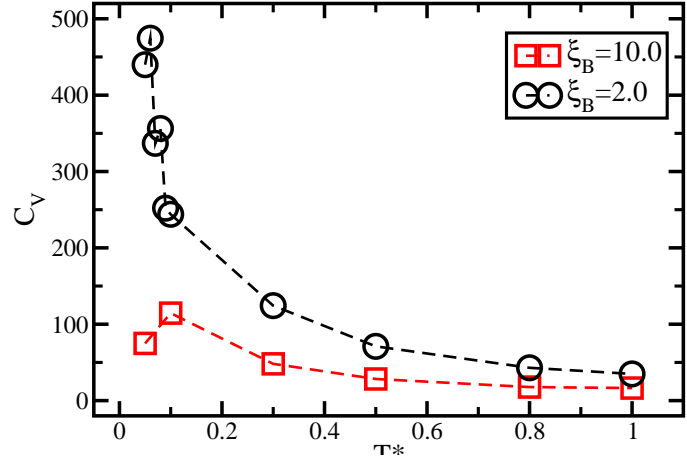}}
\subfigure[Structural relaxation time]{\includegraphics[height=6.2cm]{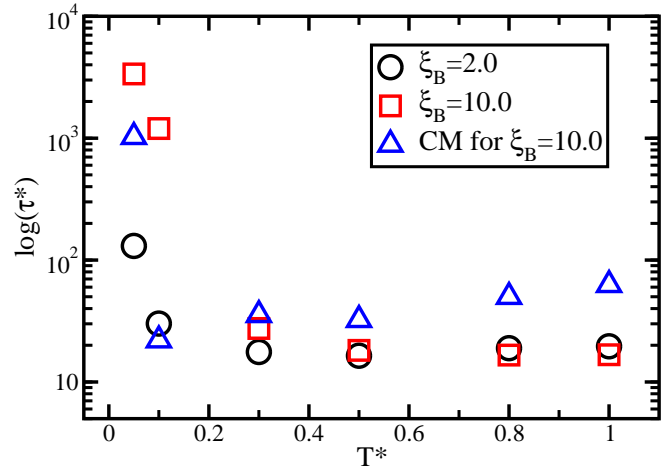}}
\caption{\label{cvtau} Specific heat, $C_V$ for all the 
temperatures $T^* = 0.05$, $0.1$, $0.3$, $0.5$, $0.8$ and $1.0$. (b) Counterion 
structural relaxation time for $\xi_b=10$ (red squares) and $\xi_B=2$ (black circle). 
The blue triangles represent polymer center-of-mass structural relaxation time 
for $2V$ system at $\xi_B=10$. At around $T^*=0.3$ the counterion $\tau^*$ becomes 
stronger than the chain CM $\tau^*$.}  
\end{figure} 

From these results presented so far, it is evident that thermodynamics of the 
charged-neutral diblock copolymers depend strongly on the electrostatic interaction 
strengths. To understand the thermodynamics better we have computed the specific 
heat capacity ($C_V$) from the fluctuation-dissipation formula, $C_V=(<E^2>-<E>^2)\big/T^{*2}$ 
where $E$ is the total energy of the system and `<>' denotes ensemble average. 
\ref{cvtau}(a) shows the $C_V$ as a function of temperature for different $\xi_B$. 
For $\xi_B=10$ (red squares), a peak in $C_V$ is observed at around $T^*=0.1$ at low 
dielectric constant. The peak is absent for $\xi_B = 2$ (black circles) at higher 
dielectric constant. Instead there is sudden jump in $C_V$ for high dielectric 
constant ($\xi_B = 2$)at around $T^*=0.1$. This suggests that the charged blocks undergo 
self-assemble (order-to-disorder transition) at this temperature. At around this temperature, 
the entropy of the system decreases thereby causing an abrupt change in $C_V$. Typically, 
the appearance of a `peak' (red squares) in $C_V$ is often referred to as the ``glass transition 
temperature". However,  the peak observed here for $\xi_B = 10$ is not associated with 
a glass transition as the system cannot vitrify at this temperature and density. At 
$\xi_B = 10$, the self-assembly is driven by electrostatic interactions rather than by entropic 
interactions. Since the $C_V$ peak occurs at low dielectric constant, it is evidently related 
to the formation of ion-counterion multiplets~\cite{Goswami1}. The multiplet 
formation 
starts occurring much 
earlier than this transition temperature as is evident from \ref{snapshot}. This phenomena 
cannot be observed at $\xi_B = 2$ as the charges are dissociated at high dielectric constant. 
At $\xi_B = 2$ the change in $C_V$ comes from the conformational energy of the chain rather than 
the multiplet formation as is evident from the abrupt increase in $C_V$ at low $T^*$. Furthermore, 
the absence of an abrupt increase in $C_V$ for $\xi_B = 10$ (low dielectric constant medium) 
can be attributed to the higher order nature of the transition that may be caused by strongly 
percolating charged blocks. We have not investigated the exact nature of this transition 
and leave this finer point for future work. The structures also break down at higher dielectric 
constant. The same phenomena has been observed in experiments~\cite{SoftMatter} where an addition 
of small trace amount of water (increasing dielectric constant) breaks down the long range order 
of the charge block copolymers. 

To understand the structural changes that generally affect the thermodynamics, we have 
computed the structural relaxation time of counterions and plotted in \ref{cvtau}(b). 
The first dynamical quantity in this article, the structural relaxation time is obtained 
from self-intermediate scattering function defined by, $S(k^*,t^*)$ where $k^*$ is 
a wave vector, $k=2\pi\big/l$ corresponds to the first peak distance ($l$) of the static 
structure factor, $S(k^*)$. The intermediate scattering function, $S(k^*,t^*$ follows 
an exponential decay. The structural relaxation time, $\tau^*$ is represented by the 
value of $t^*$ where $S(k^*,t^*)$ resumes $1/e$ times initial amplitude. The red squares 
and blue triangles are shown for counterions and chain center-of-mass (CM) respectively 
for $\xi_B = 10$, i.e., at low dielectric constant and the black circles are shown for 
counterions at $\xi_B = 2$. For $\xi_B = 2$, the chain CM relaxation time is not shown 
as the major focus of this investigation is the effect of counterion relaxation. At 
low dielectric constant, formation of charge multiplets does not allow the counterions 
relax thereby a slow relaxation can be observed (red squares). These plots show an 
abrupt change in $\tau^*$ at low temperature which is consistent with the $C_V$ 
plot shown in \ref{cvtau}(a). For the low dielectric constant ($\xi_B=10$), the 
counterions cross the chain CM at around $T^*=0.1$, exactly the temperature at which 
the peak in $C_V$ has been observed. These results support the theory that the counterions 
undergo a structural transition at low temperatures due to multiplet formation. Below this 
temperature the counterion relaxation time increases strongly, consistent with the 
localization of charged blocks into multiplet clusters. A similar behavior has been 
observed in other weakly charged systems where the transition occurs due to the 
formation of charge multiplets~\cite{Goswami1}.  

\begin{figure}[tbp]
\pagebreak 
\centering
\includegraphics[height=6.2cm]{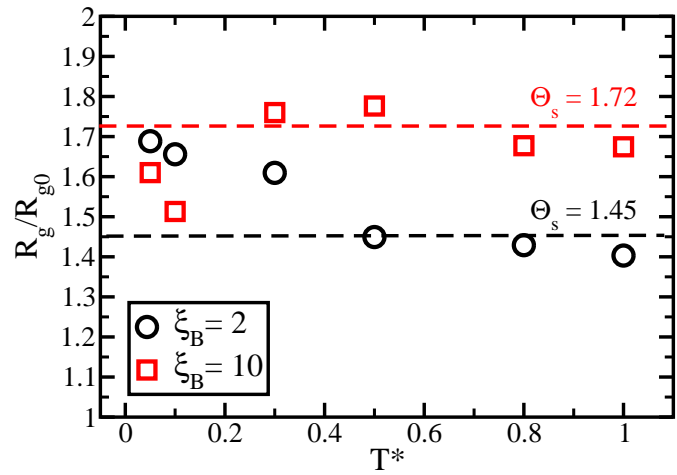}
\caption{\label{rgratio} Ratio of radius-of-gyration to the radius-of-gyration of an ideal chain 
of same chain length for $\xi_B=2$ (black circles) and $\xi_B=10$ (red squares) systems. For 
$\xi_B=10$, system size $2V$ is used. $\xi_B = 10$ shows more chain swelling compared 
to $\xi_B=2$. The lines are not data, these are merely guide to the eyes for the respective 
swelling parameters.} 
\end{figure} 

To understand the swelling behavior of the charged diblock copolymer for different 
dielectric constants, we compare the radius of gyration of the charged polymer ($R_g$) with 
that of an ideal chain, $R_{g0}=\sqrt{Nb^2\big/6}$, where $N$ is the degree of polymerization 
and $b$ is the monomer length. The swelling ratio, $\Theta_s = R_g\big/R_{g0}$ versus 
temperature is shown in \ref{rgratio}. The ratio, $\Theta_s$ is similar around 
the transition temperature for both $\xi_B = 2$ (black circle) and $10$ (red squares). 
However, beyond the transition temperature, $\Theta_s$ is slightly higher for $\xi_B=10$ 
than $\xi_B = 2$. This can be explained as follows. As the dielectric constant 
increases ($\xi_B = 2$), swelling capability of the chain decreases which 
can normally happen in neutral polymers only. Therefore, the swelling phenomena of 
the charged BCP system for $\xi_B = 2$ may be attributed to the swelling of 
neutral chains. On the other hand, for polyelectrolytes, increase in dielectric constant 
increases swelling capability. Therefore for $\xi_B = 10$ (red squares), increase 
in swelling phenomena can be attributed to the polyelectrolyte behavior of the charged BCP. 
In these data comparatively small swelling of the chain has been observed, in some cases 
though, the swelling has been observed to be as much as 100 times their volume for particular 
classes of polyelectrolytes in low-polarity (low-dielectric constant) solvents~\cite{Ono1}. 
This suggests that, although there are dissociation of counterions to break the 
multiplets with the increase in dielectric constant, the entropy due to the 
dissociation of counterions is relatively small to contribute substantially in 
swelling of the charged BCP.  The system under consideration has only 8 
charges (and 8 counterions) on the first 16 monomers of a chain of length 64. 
These small number of charges result in statistically smaller contribution to 
the entropy of dissociation, although the Coulombic interactions significantly 
contribute to the formation and breaking of charge agglomeration as shown in 
\ref{snapshot} and \ref{snapshot_xi2}. 

\begin{figure}[tbp]
\pagebreak 
\centering
\subfigure[Counterion MSD]{\includegraphics[height=6.2cm]{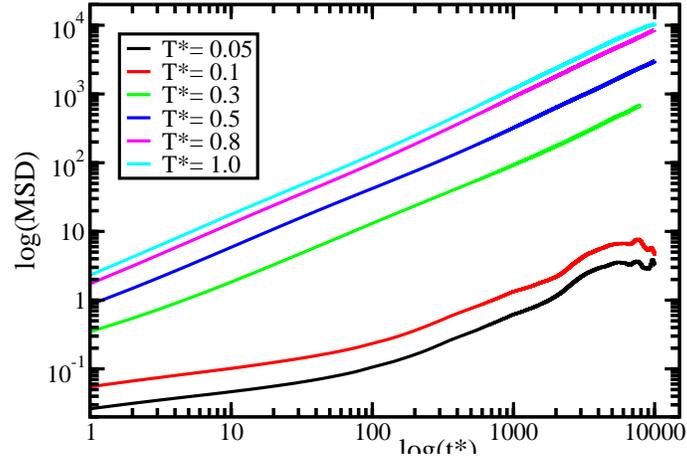}}
\subfigure[Chain CM MSD]{\includegraphics[height=6.2cm]{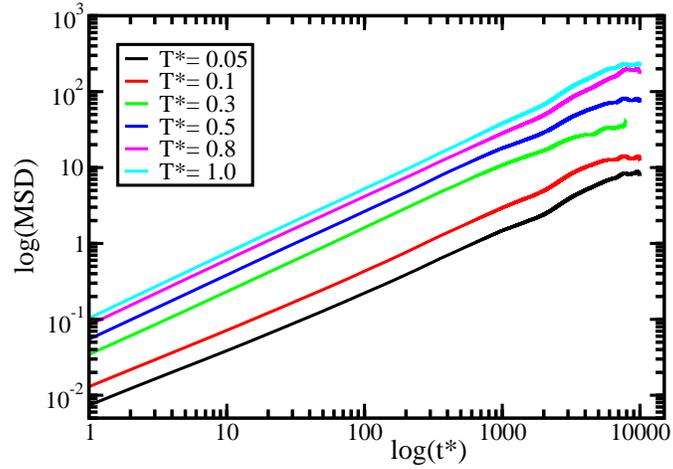}}
\caption{\label{msdV} Mean square displacement at $\xi_B=10.0$ for (a) 
counterions and (b) chain center-of-mass.  Black line represent 
$T^* = 0.05$, red line $0.1$, green line $0.3$, blue line $0.5$, 
magenta line $0.8$ and cyan line $1.0$ respectively. MSD is shown for 
$V$ system due to lack of data for $2\times V$ system.} 
\end{figure}

\begin{figure}[tbp]
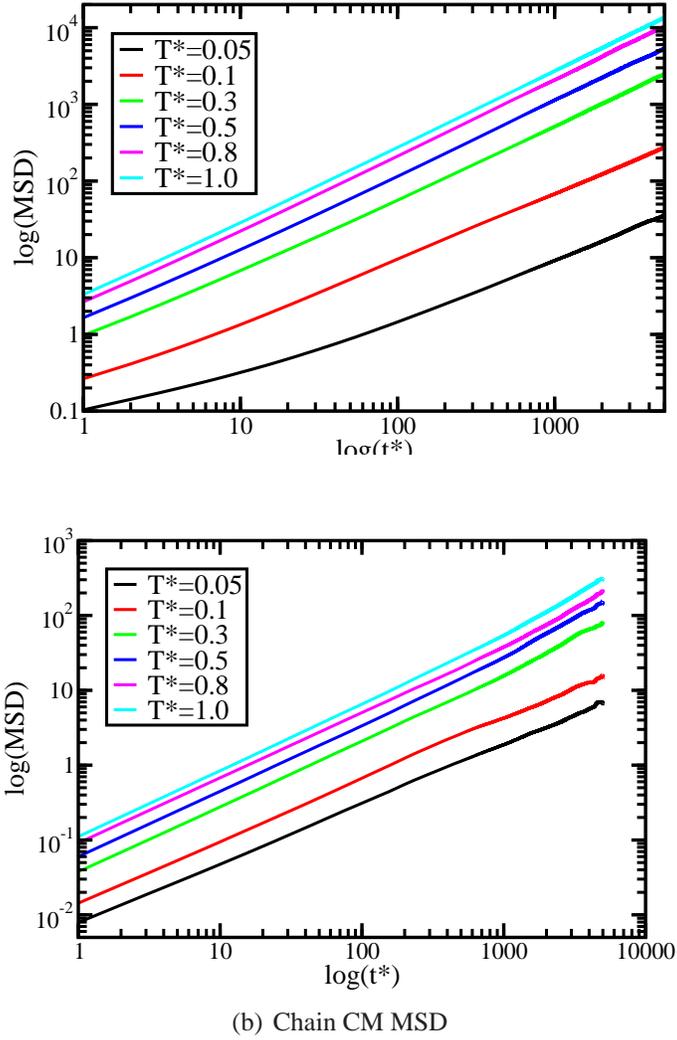

\pagebreak 
\centering
\subfigure[Counterion MSD]{\includegraphics[height=6.2cm]{figure7a.eps}}
\subfigure[Chain CM MSD]{\includegraphics[height=6.2cm]{figure7b.eps}}
\caption{\label{msdXi2} Mean square displacement at $\xi_B=2.0$ for (a) 
counterions and (b) chain center-of-mass.  Respective temperatures for 
the lines are shown in the legends.}  
\end{figure} 

\ref{msdV} and \ref{msdXi2} show the mean-square-displacement (MSD) of the 
counterions and chain CM for two different $\xi_B$. For $\xi_B=10$ (\ref{msdV}), 
only the $V$ system is shown due to lack of data to calculate correlation 
in the $2V$ system, but this should not be a major concern as there is hardly 
any system size effects present in these calculations. In \ref{msdV}(a) 
and \ref{msdXi2} counterion motion is observed to be much faster than the 
chain CM motion (shown in \ref{msdV}(b) and \ref{msdXi2}(b)). At low 
temperatures, both 
counterions and chain CM show sharp decrease in their respective MSDs. Although
counterions show faster motion due to their smaller size compared to the chain CM, 
the drop in counterion MSD is more pronounced than the chain CM at 
$T^* = 0.05$ and $0.1$ respectively. 
This may be due to the stronger 
agglomeration of counterions that causes the counterions to be entropically 
unfavorable to move. For low dielectric constant, $\xi_B=10$ (\ref{msdV}(a) the 
low temperature counterion motions become even more sluggish. This is consistent 
with the earlier observation of counterions forming charge multiplets at 
low temperatures for high $\xi_B$. For $\xi_B=2$, the counterions dissociate 
due to the presence of more polar medium (high dielectric constant) therefore 
the motion is enhanced, although slow compared to the chain CM. The sudden 
jump in MSD for both the chain CM and counterions for both $\xi_B =2$ and 
$10$ suggests a transition from the non-agglomerated state to charge 
agglomerated state.  

\section{Conclusion} 
\label{conclusion} 
In summary, we have simulated a system of charged block copolymers in the 
presence of explicit counterions. The partially charged block copolymers 
exhibit `inverse' morphologies that have been observed in recent experiments~\cite{SoftMatter}. 
From our results, it is evident that the design of these novel functional materials depend 
sensitively upon the dielectric constant of the medium, the degree of ionization and the 
temperature of the system. Corroborative results have been obtained 
experimentally~\cite{SoftMatter} that addition of trace amount of water (increasing 
dielectric constant) changes the morphology of the diblock copolymer dramatically.  

Furthermore, our results highlight the existence of two regimes in the self-assembly 
of charged-neutral diblock copolymers. In one regime, the self-assembly is governed 
by electrostatic energy and in the other regime, it is dictated by the counterion entropy 
affecting the morphologies and other thermodynamic properties. It is interesting to note 
that introducing charges on a diblock copolymer not only changes the morphologies, it 
acts unfavorably for both neutral polymers and polyelectrolytes depending on the dielectric 
of the medium.  In a low dielectric medium, the self-assembly in charged-neutral diblock 
is electrostatically dominated and leads to non-trivial, counter-intuitive `inverse' 
morphologies where the minority component forms the matrix and majority forms the 
structures. For this system, electrostatics drive the formation of ionic multiplets which 
increases the relaxation time of the counterions and hence affect the dynamics of microphase 
separation (\ref{snapshot}). For high dielectric medium, the chains form neutral 
BCP morphologies as can be seen in \ref{snapshot_xi2}. An increase in dielectric 
constant shifts the system towards counterion entropy dominated regime. The counterion 
dominated regime also can be observed with increase in temperature. In this regime, 
disordered phase gets stabilized against ordered structures. These results exhibit 
how different the 75-25 charged BCPs are from the neutral diblocks counterparts. 
The swelling 
behavior 
shown in \ref{rgratio} shows how different these charged BCPs are from their 
polyelectrolyte counterparts. The swelling of the chains, although very small, shows 
behavior which is opposite to polyelectrolytes swelling behavior at low $D$. 
The formation of charge multiplets in a cluster and the dissociation of 
counterions with decreasing dielectric constant, can be understood by 
observing the dynamics of counterions (\ref{msdV} and \ref{msdXi2}). 
These results are in agreement with the SCFT calculations~\cite{kumar} for weakly 
charged systems, showing the stabilization of disorder phase against 
ordered morphologies to avoid the entropic cost of partitioning of 
counterions in the charged domains. 
It should be noted here that a discussion on the 
slope of the MSD (diffusion coefficients) is missing. A detailed discussion 
of the diffusion coefficients of the chain CM and counterions may indeed be 
important to understand the effect of transport coefficients on the 
morphological changes for different $T^*$ and $D$, however, diffusion 
in polymers is a subject that deserves a complete separate study. 

The present structural and dynamical analysis of charged diblock polymers should be 
of importance for understanding the fundamental mechanisms and for designing 
novel materials for drug delivery~\cite{Gaucher1,Garrec1,Cui1,Guo2,Guo1} 
and nanotechnology~\cite{Lodge1,Lodge2} applications. Additional information of 
different charged states and block sizes of the charged block will further enhance 
the understanding and in particular provide information for counterion transport in 
charged polymer films.  

\acknowledgement 
This work was supported by the Division of Materials Science and 
Engineering (DMSE), U.S. Department of Energy (DoE), Office of Basic 
Energy Sciences (BES) under Contract No. DEAC05-00OR22725 with 
UT-Battelle, LLC at Oak Ridge National Laboratory (ORNL).

\bibliography{references} 

\providecommand*\mcitethebibliography{\thebibliography}
\csname @ifundefined\endcsname{endmcitethebibliography}
  {\let\endmcitethebibliography\endthebibliography}{}
\begin{mcitethebibliography}{58}
\providecommand*\natexlab[1]{#1}
\providecommand*\mciteSetBstSublistMode[1]{}
\providecommand*\mciteSetBstMaxWidthForm[2]{}
\providecommand*\mciteBstWouldAddEndPuncttrue
  {\def\EndOfBibitem{\unskip.}}
\providecommand*\mciteBstWouldAddEndPunctfalse
  {\let\EndOfBibitem\relax}
\providecommand*\mciteSetBstMidEndSepPunct[3]{}
\providecommand*\mciteSetBstSublistLabelBeginEnd[3]{}
\providecommand*\EndOfBibitem{}
\mciteSetBstSublistMode{f}
\mciteSetBstMaxWidthForm{subitem}{(\alph{mcitesubitemcount})}
\mciteSetBstSublistLabelBeginEnd
  {\mcitemaxwidthsubitemform\space}
  {\relax}
  {\relax}

\bibitem[Discher et~al.(1999)Discher, Won, Ege, Lee, Bates, Discher, and
  Hammer]{Bates1}
Discher,~B.~M.; Won,~Y.-Y.; Ege,~D.~S.; Lee,~C.-M.; Bates,~F.~S.;
  Discher,~D.~E.; Hammer,~D.~A. \emph{Science} \textbf{1999}, \emph{284},
  1143\relax
\mciteBstWouldAddEndPuncttrue
\mciteSetBstMidEndSepPunct{\mcitedefaultmidpunct}
{\mcitedefaultendpunct}{\mcitedefaultseppunct}\relax
\EndOfBibitem
\bibitem[Bates and Fredrickson(1990)Bates, and Fredrickson]{Bates2}
Bates,~F.~S.; Fredrickson,~G.~H. \emph{Annu.~Rev.~Phys.~Chem.} \textbf{1990},
  \emph{41}, 525\relax
\mciteBstWouldAddEndPuncttrue
\mciteSetBstMidEndSepPunct{\mcitedefaultmidpunct}
{\mcitedefaultendpunct}{\mcitedefaultseppunct}\relax
\EndOfBibitem
\bibitem[Russell(1996)]{Russell1}
Russell,~T.~P. \emph{Curr. Opin. Colloid Inerface Sci.} \textbf{1996},
  \emph{1}, 107\relax
\mciteBstWouldAddEndPuncttrue
\mciteSetBstMidEndSepPunct{\mcitedefaultmidpunct}
{\mcitedefaultendpunct}{\mcitedefaultseppunct}\relax
\EndOfBibitem
\bibitem[Bates and Fredrickson(1999)Bates, and Fredrickson]{Bates3}
Bates,~F.~S.; Fredrickson,~G.~H. \emph{Phys. Today} \textbf{1999}, \emph{52},
  32\relax
\mciteBstWouldAddEndPuncttrue
\mciteSetBstMidEndSepPunct{\mcitedefaultmidpunct}
{\mcitedefaultendpunct}{\mcitedefaultseppunct}\relax
\EndOfBibitem
\bibitem[Fasolka and Mayes(2001)Fasolka, and Mayes]{Fasolka1}
Fasolka,~M.~J.; Mayes,~A.~M. \emph{Annu.~Rev.~Mater.~Res.} \textbf{2001},
  \emph{11}, 323\relax
\mciteBstWouldAddEndPuncttrue
\mciteSetBstMidEndSepPunct{\mcitedefaultmidpunct}
{\mcitedefaultendpunct}{\mcitedefaultseppunct}\relax
\EndOfBibitem
\bibitem[Vriezema et~al.(2005)Vriezema, Aragones, Elemans, Cornelissen, Rowan,
  and Nolte]{Vriezema1}
Vriezema,~D.~M.; Aragones,~M.~C.; Elemans,~J.~A.~A.~W.;
  Cornelissen,~J.~J.~L.~M.; Rowan,~A.~E.; Nolte,~R.~J.~M. \emph{Chem.~Rev.}
  \textbf{2005}, \emph{105}, 1445\relax
\mciteBstWouldAddEndPuncttrue
\mciteSetBstMidEndSepPunct{\mcitedefaultmidpunct}
{\mcitedefaultendpunct}{\mcitedefaultseppunct}\relax
\EndOfBibitem
\bibitem[Zhang et~al.(2009)Zhang, Rehm, Safont-Sempere, and Wurthner]{Zhang1}
Zhang,~X.; Rehm,~S.; Safont-Sempere,~M.~M.; Wurthner,~F. \emph{Nat.~Chem.}
  \textbf{2009}, \emph{1}, 623\relax
\mciteBstWouldAddEndPuncttrue
\mciteSetBstMidEndSepPunct{\mcitedefaultmidpunct}
{\mcitedefaultendpunct}{\mcitedefaultseppunct}\relax
\EndOfBibitem
\bibitem[Jones(2009)]{Jones1}
Jones,~R.~A.~L. \emph{Faraday ~Discuss.} \textbf{2009}, \emph{143}, 9\relax
\mciteBstWouldAddEndPuncttrue
\mciteSetBstMidEndSepPunct{\mcitedefaultmidpunct}
{\mcitedefaultendpunct}{\mcitedefaultseppunct}\relax
\EndOfBibitem
\bibitem[Kim et~al.(2009)Kim, Cornelissen, and van Hest]{Kim1}
Kim,~K.~T.; Cornelissen,~J.~J.~L.~M.; van Hest,~R.~J.~M.~N. J.~C.~M.
  \emph{Adv.~Mater.} \textbf{2009}, \emph{21}, 2787\relax
\mciteBstWouldAddEndPuncttrue
\mciteSetBstMidEndSepPunct{\mcitedefaultmidpunct}
{\mcitedefaultendpunct}{\mcitedefaultseppunct}\relax
\EndOfBibitem
\bibitem[Ariga et~al.(2008)Ariga, Hill, Lee, Vinu, Charvet, and
  Acharya]{Ariga1}
Ariga,~K.; Hill,~J.~P.; Lee,~M.~V.; Vinu,~A.; Charvet,~R.; Acharya,~S.
  \emph{Sci. Technol. Adv. Mater.} \textbf{2008}, \emph{9}, 014109\relax
\mciteBstWouldAddEndPuncttrue
\mciteSetBstMidEndSepPunct{\mcitedefaultmidpunct}
{\mcitedefaultendpunct}{\mcitedefaultseppunct}\relax
\EndOfBibitem
\bibitem[Li et~al.(2007)Li, Chen, Chu, Hales, Wooley, and Pochan]{Li1}
Li,~Z.; Chen,~Z.; Chu,~H.; Hales,~K.; Wooley,~K.~L.; Pochan,~D.~J.
  \emph{Langmuir} \textbf{2007}, \emph{23}, 4689\relax
\mciteBstWouldAddEndPuncttrue
\mciteSetBstMidEndSepPunct{\mcitedefaultmidpunct}
{\mcitedefaultendpunct}{\mcitedefaultseppunct}\relax
\EndOfBibitem
\bibitem[Zhang and Eisenberg(1995)Zhang, and Eisenberg]{Eisenberg1}
Zhang,~L.; Eisenberg,~A. \emph{Science} \textbf{1995}, \emph{268}, 1728\relax
\mciteBstWouldAddEndPuncttrue
\mciteSetBstMidEndSepPunct{\mcitedefaultmidpunct}
{\mcitedefaultendpunct}{\mcitedefaultseppunct}\relax
\EndOfBibitem
\bibitem[Gaucher et~al.(2005)Gaucher, Dufresne, Sant, Kang, Maysinger, and
  Leroux]{Gaucher1}
Gaucher,~G.; Dufresne,~M.~H.; Sant,~V.~P.; Kang,~N.; Maysinger,~D.;
  Leroux,~J.~C. \emph{J.~Controlled Release} \textbf{2005}, \emph{109},
  169\relax
\mciteBstWouldAddEndPuncttrue
\mciteSetBstMidEndSepPunct{\mcitedefaultmidpunct}
{\mcitedefaultendpunct}{\mcitedefaultseppunct}\relax
\EndOfBibitem
\bibitem[Garrec et~al.(2004)Garrec, Ranger, and Leroux]{Garrec1}
Garrec,~D.~L.; Ranger,~M.; Leroux,~J.~C. \emph{Am.~J.~Drug Delivery}
  \textbf{2004}, \emph{2}, 15\relax
\mciteBstWouldAddEndPuncttrue
\mciteSetBstMidEndSepPunct{\mcitedefaultmidpunct}
{\mcitedefaultendpunct}{\mcitedefaultseppunct}\relax
\EndOfBibitem
\bibitem[Hu and Jing(2009)Hu, and Jing]{Hu1}
Hu,~X.~L.; Jing,~X.~B. \emph{Expert Opin. Drug Discovery} \textbf{2009},
  \emph{6}, 1079\relax
\mciteBstWouldAddEndPuncttrue
\mciteSetBstMidEndSepPunct{\mcitedefaultmidpunct}
{\mcitedefaultendpunct}{\mcitedefaultseppunct}\relax
\EndOfBibitem
\bibitem[Savic et~al.(2009)Savic, Azzam, Eisenberg, Nedev, Rosenberg, and
  Maysinger]{Savic1}
Savic,~R.; Azzam,~T.; Eisenberg,~A.; Nedev,~H.; Rosenberg,~L.; Maysinger,~D.
  \emph{Biomaterials} \textbf{2009}, \emph{30}, 3597\relax
\mciteBstWouldAddEndPuncttrue
\mciteSetBstMidEndSepPunct{\mcitedefaultmidpunct}
{\mcitedefaultendpunct}{\mcitedefaultseppunct}\relax
\EndOfBibitem
\bibitem[Bajpai et~al.(2008)Bajpai, Shukla, Bhanu, and Kankane]{Bajpai1}
Bajpai,~A.~K.; Shukla,~S.~K.; Bhanu,~S.; Kankane,~S. \emph{Prog.~Polym.~Sci.}
  \textbf{2008}, \emph{33}, 1088\relax
\mciteBstWouldAddEndPuncttrue
\mciteSetBstMidEndSepPunct{\mcitedefaultmidpunct}
{\mcitedefaultendpunct}{\mcitedefaultseppunct}\relax
\EndOfBibitem
\bibitem[Beduneau et~al.(2007)Beduneau, Saulnier, and Benoit]{Beduneau1}
Beduneau,~A.; Saulnier,~P.; Benoit,~J.~P. \emph{Biomaterials} \textbf{2007},
  \emph{28}, 4947\relax
\mciteBstWouldAddEndPuncttrue
\mciteSetBstMidEndSepPunct{\mcitedefaultmidpunct}
{\mcitedefaultendpunct}{\mcitedefaultseppunct}\relax
\EndOfBibitem
\bibitem[Jain and Bates(2004)Jain, and Bates]{Jain1}
Jain,~S.; Bates,~F.~S. \emph{Macromolecules} \textbf{2004}, \emph{37},
  1511\relax
\mciteBstWouldAddEndPuncttrue
\mciteSetBstMidEndSepPunct{\mcitedefaultmidpunct}
{\mcitedefaultendpunct}{\mcitedefaultseppunct}\relax
\EndOfBibitem
\bibitem[Halperin et~al.(1992)Halperin, Tirrell, and Lodge]{Halperin2}
Halperin,~A.; Tirrell,~M.; Lodge,~T.~P. \emph{Adv.~Polym.~Sci.} \textbf{1992},
  \emph{100}, 31\relax
\mciteBstWouldAddEndPuncttrue
\mciteSetBstMidEndSepPunct{\mcitedefaultmidpunct}
{\mcitedefaultendpunct}{\mcitedefaultseppunct}\relax
\EndOfBibitem
\bibitem[Cui et~al.(2007)Cui, Chen, Chong, Wooley, and Pochan]{Cui1}
Cui,~H.; Chen,~Z.; Chong,~S.; Wooley,~K.~L.; Pochan,~D.~J. \emph{Science}
  \textbf{2007}, \emph{317}, 647\relax
\mciteBstWouldAddEndPuncttrue
\mciteSetBstMidEndSepPunct{\mcitedefaultmidpunct}
{\mcitedefaultendpunct}{\mcitedefaultseppunct}\relax
\EndOfBibitem
\bibitem[Li et~al.(2006)Li, Tang, Qiu, Zhang, and Yang]{Li3}
Li,~X.~A.; Tang,~P.; Qiu,~F.; Zhang,~H.~D.; Yang,~Y.~L. \emph{J.~Phys.~Chem B}
  \textbf{2006}, \emph{110}, 2024\relax
\mciteBstWouldAddEndPuncttrue
\mciteSetBstMidEndSepPunct{\mcitedefaultmidpunct}
{\mcitedefaultendpunct}{\mcitedefaultseppunct}\relax
\EndOfBibitem
\bibitem[Ortiz et~al.(2005)Ortiz, Nielson, Discher, Klein, Lipowsky, and
  Shillcock]{Ortiz1}
Ortiz,~V.; Nielson,~D.~E.; Discher,~D.~E.; Klein,~M.~I.; Lipowsky,~R.;
  Shillcock,~J. \emph{J.~Phys.~Chem B} \textbf{2005}, \emph{109}, 17708\relax
\mciteBstWouldAddEndPuncttrue
\mciteSetBstMidEndSepPunct{\mcitedefaultmidpunct}
{\mcitedefaultendpunct}{\mcitedefaultseppunct}\relax
\EndOfBibitem
\bibitem[Guo et~al.(2009)Guo, Tan, Kim, Zhang, Zhang, Hedrick, Yang, and
  Qian]{Guo2}
Guo,~X.~D.; Tan,~J.~P.~K.; Kim,~S.~H.; Zhang,~L.~J.; Zhang,~Y.; Hedrick,~J.~L.;
  Yang,~Y.~Y.; Qian,~Y. \emph{Biomaterials} \textbf{2009}, \emph{30},
  6556\relax
\mciteBstWouldAddEndPuncttrue
\mciteSetBstMidEndSepPunct{\mcitedefaultmidpunct}
{\mcitedefaultendpunct}{\mcitedefaultseppunct}\relax
\EndOfBibitem
\bibitem[Guo et~al.(2009)Guo, Tan, Zhang, Khan, Liu, Yang, and Qian]{Guo1}
Guo,~X.~D.; Tan,~J.~P.~K.; Zhang,~L.~J.; Khan,~M.; Liu,~S.~Q.; Yang,~Y.~Y.;
  Qian,~Y. \emph{Chem.~Phys.~Lett.} \textbf{2009}, \emph{473}, 336\relax
\mciteBstWouldAddEndPuncttrue
\mciteSetBstMidEndSepPunct{\mcitedefaultmidpunct}
{\mcitedefaultendpunct}{\mcitedefaultseppunct}\relax
\EndOfBibitem
\bibitem[Sknepnek et~al.(2008)Sknepnek, Anderson, Lamm, Schmalian, and
  Travesset]{Sknepnek1}
Sknepnek,~R.; Anderson,~J.~A.; Lamm,~M.~H.; Schmalian,~J.~R.; Travesset,~A.
  \emph{ACS Nano} \textbf{2008}, \emph{2}, 1259\relax
\mciteBstWouldAddEndPuncttrue
\mciteSetBstMidEndSepPunct{\mcitedefaultmidpunct}
{\mcitedefaultendpunct}{\mcitedefaultseppunct}\relax
\EndOfBibitem
\bibitem[Erdtman et~al.(2008)Erdtman, dos Santos, Lofgren, and
  Eriksson]{Erdtman1}
Erdtman,~E.; dos Santos,~D.~J.~V.~A.; Lofgren,~L.; Eriksson,~L.~A.
  \emph{Chem.~Phys.~Lett} \textbf{2008}, \emph{463}, 178\relax
\mciteBstWouldAddEndPuncttrue
\mciteSetBstMidEndSepPunct{\mcitedefaultmidpunct}
{\mcitedefaultendpunct}{\mcitedefaultseppunct}\relax
\EndOfBibitem
\bibitem[Srinivas et~al.(2004)Srinivas, Discher, and Klein]{Srinivas1}
Srinivas,~G.; Discher,~D.~E.; Klein,~M.~I. \emph{Nat.~Matter.} \textbf{2004},
  \emph{3}, 638\relax
\mciteBstWouldAddEndPuncttrue
\mciteSetBstMidEndSepPunct{\mcitedefaultmidpunct}
{\mcitedefaultendpunct}{\mcitedefaultseppunct}\relax
\EndOfBibitem
\bibitem[Goswami et~al.(2010)Goswami, Sumpter, and Mays]{ChemPhysLett}
Goswami,~M.; Sumpter,~B.~G.; Mays,~J.~W. \emph{Chem.~Phys.~Lett} \textbf{2010},
  \emph{487}, 272\relax
\mciteBstWouldAddEndPuncttrue
\mciteSetBstMidEndSepPunct{\mcitedefaultmidpunct}
{\mcitedefaultendpunct}{\mcitedefaultseppunct}\relax
\EndOfBibitem
\bibitem[Deschenes et~al.(2008)Deschenes, Bousmina, and Ritcey]{Deschenes1}
Deschenes,~I.; Bousmina,~M.; Ritcey,~A.~M. \emph{Langmuir} \textbf{2008},
  \emph{24}, 3699\relax
\mciteBstWouldAddEndPuncttrue
\mciteSetBstMidEndSepPunct{\mcitedefaultmidpunct}
{\mcitedefaultendpunct}{\mcitedefaultseppunct}\relax
\EndOfBibitem
\bibitem[Du et~al.(2007)Du, Zhu, and Jiang]{Du1}
Du,~H.~B.; Zhu,~J.~T.; Jiang,~W. \emph{J.~Phys.~Chem B} \textbf{2007},
  \emph{111}, 1938\relax
\mciteBstWouldAddEndPuncttrue
\mciteSetBstMidEndSepPunct{\mcitedefaultmidpunct}
{\mcitedefaultendpunct}{\mcitedefaultseppunct}\relax
\EndOfBibitem
\bibitem[Ma et~al.(2007)Ma, Li, Tang, and Yang]{Ma1}
Ma,~J.~W.; Li,~X.; Tang,~P.; Yang,~Y.~I. \emph{J.~Phys.~Chem B} \textbf{2007},
  \emph{111}, 1552\relax
\mciteBstWouldAddEndPuncttrue
\mciteSetBstMidEndSepPunct{\mcitedefaultmidpunct}
{\mcitedefaultendpunct}{\mcitedefaultseppunct}\relax
\EndOfBibitem
\bibitem[Sevinek and Zvelindovsky(2005)Sevinek, and Zvelindovsky]{Sevink1}
Sevinek,~G.~J.~A.; Zvelindovsky,~A.~V. \emph{Macromolecules} \textbf{2005},
  \emph{38}, 7502\relax
\mciteBstWouldAddEndPuncttrue
\mciteSetBstMidEndSepPunct{\mcitedefaultmidpunct}
{\mcitedefaultendpunct}{\mcitedefaultseppunct}\relax
\EndOfBibitem
\bibitem[Li et~al.(2009)Li, Guo, Liu, and Liang]{Li}
Li,~X.; Guo,~J.; Liu,~Y.; Liang,~H. \emph{J.~Chem.~Phys.} \textbf{2009},
  \emph{130}, 074908\relax
\mciteBstWouldAddEndPuncttrue
\mciteSetBstMidEndSepPunct{\mcitedefaultmidpunct}
{\mcitedefaultendpunct}{\mcitedefaultseppunct}\relax
\EndOfBibitem
\bibitem[Kriksin et~al.(2009)Kriksin, Khalatur, ten Brinke, and
  Khokhlov]{Kriksin1}
Kriksin,~Y.~A.; Khalatur,~P.~G.; ten Brinke,~I.~Y.~E.~G.; Khokhlov,~A.~R.
  \emph{Soft Matter} \textbf{2009}, \emph{5}, 2896\relax
\mciteBstWouldAddEndPuncttrue
\mciteSetBstMidEndSepPunct{\mcitedefaultmidpunct}
{\mcitedefaultendpunct}{\mcitedefaultseppunct}\relax
\EndOfBibitem
\bibitem[Khokhlov and Khalatur(2008)Khokhlov, and Khalatur]{Khokhlov1}
Khokhlov,~A.~R.; Khalatur,~P.~G. \emph{Chem.~Phys.~Lett} \textbf{2008},
  \emph{461}, 58\relax
\mciteBstWouldAddEndPuncttrue
\mciteSetBstMidEndSepPunct{\mcitedefaultmidpunct}
{\mcitedefaultendpunct}{\mcitedefaultseppunct}\relax
\EndOfBibitem
\bibitem[Xin et~al.(2009)Xin, Liu, and Zhong]{Xin1}
Xin,~J.; Liu,~D.; Zhong,~C. \emph{J.~Phys.~Chem.~B} \textbf{2009}, \emph{113},
  9364\relax
\mciteBstWouldAddEndPuncttrue
\mciteSetBstMidEndSepPunct{\mcitedefaultmidpunct}
{\mcitedefaultendpunct}{\mcitedefaultseppunct}\relax
\EndOfBibitem
\bibitem[Marko and Rabin(1991)Marko, and Rabin]{marko1}
Marko,~J.; Rabin,~Y. \emph{Macromolecules} \textbf{1991}, \emph{24}, 2134\relax
\mciteBstWouldAddEndPuncttrue
\mciteSetBstMidEndSepPunct{\mcitedefaultmidpunct}
{\mcitedefaultendpunct}{\mcitedefaultseppunct}\relax
\EndOfBibitem
\bibitem[Marko and Rabin(1992)Marko, and Rabin]{marko2}
Marko,~J.; Rabin,~Y. \emph{Macromolecules} \textbf{1992}, \emph{25}, 1503\relax
\mciteBstWouldAddEndPuncttrue
\mciteSetBstMidEndSepPunct{\mcitedefaultmidpunct}
{\mcitedefaultendpunct}{\mcitedefaultseppunct}\relax
\EndOfBibitem
\bibitem[Kumar and Muthukumar(2007)Kumar, and Muthukumar]{kumar}
Kumar,~R.; Muthukumar,~M. \emph{J.~Chem. ~Phys} \textbf{2007}, \emph{126},
  214902\relax
\mciteBstWouldAddEndPuncttrue
\mciteSetBstMidEndSepPunct{\mcitedefaultmidpunct}
{\mcitedefaultendpunct}{\mcitedefaultseppunct}\relax
\EndOfBibitem
\bibitem[Park and Balsara(2008)Park, and Balsara]{balsara}
Park,~M.; Balsara,~N. \emph{Macromolecules} \textbf{2008}, \emph{41},
  3678\relax
\mciteBstWouldAddEndPuncttrue
\mciteSetBstMidEndSepPunct{\mcitedefaultmidpunct}
{\mcitedefaultendpunct}{\mcitedefaultseppunct}\relax
\EndOfBibitem
\bibitem[Manning(1969)]{manning}
Manning,~G.~S. \emph{J.\ Chem.\ Phys.} \textbf{1969}, \emph{51}, 924\relax
\mciteBstWouldAddEndPuncttrue
\mciteSetBstMidEndSepPunct{\mcitedefaultmidpunct}
{\mcitedefaultendpunct}{\mcitedefaultseppunct}\relax
\EndOfBibitem
\bibitem[Liu and Muthukumar(2002)Liu, and Muthukumar]{muthu_shulan}
Liu,~S.; Muthukumar,~M. \emph{J.\ Chem.\ Phys.} \textbf{2002}, \emph{116},
  9975\relax
\mciteBstWouldAddEndPuncttrue
\mciteSetBstMidEndSepPunct{\mcitedefaultmidpunct}
{\mcitedefaultendpunct}{\mcitedefaultseppunct}\relax
\EndOfBibitem
\bibitem[Muthukumar(2004)]{muthu_count}
Muthukumar,~M. \emph{J.\ Chem.\ Phys.} \textbf{2004}, \emph{120}, 9343\relax
\mciteBstWouldAddEndPuncttrue
\mciteSetBstMidEndSepPunct{\mcitedefaultmidpunct}
{\mcitedefaultendpunct}{\mcitedefaultseppunct}\relax
\EndOfBibitem
\bibitem[Kumar et~al.(2009)Kumar, Kundagrami, and Muthukumar]{muthu_kumar}
Kumar,~R.; Kundagrami,~A.; Muthukumar,~M. \emph{Macromolecules} \textbf{2009},
  \emph{42}, 1370\relax
\mciteBstWouldAddEndPuncttrue
\mciteSetBstMidEndSepPunct{\mcitedefaultmidpunct}
{\mcitedefaultendpunct}{\mcitedefaultseppunct}\relax
\EndOfBibitem
\bibitem[Goswami et~al.(2010)Goswami, Sumpter, Huang, Messman, Gido,
  Isaacs-Sodeye, and Mays]{SoftMatter}
Goswami,~M.; Sumpter,~B.~G.; Huang,~T.; Messman,~J.~M.; Gido,~S.~P.;
  Isaacs-Sodeye,~A.~I.; Mays,~J.~W. \emph{Soft Matter} \textbf{2010},
  \emph{DOI: 10.1039/c0sm00733a}\relax
\mciteBstWouldAddEndPuncttrue
\mciteSetBstMidEndSepPunct{\mcitedefaultmidpunct}
{\mcitedefaultendpunct}{\mcitedefaultseppunct}\relax
\EndOfBibitem
\bibitem[Kremer and Grest(1990)Kremer, and Grest]{kremergrest2}
Kremer,~K.; Grest,~G.~S. \emph{J.\ Chem.\ Phys.} \textbf{1990}, \emph{92},
  5057\relax
\mciteBstWouldAddEndPuncttrue
\mciteSetBstMidEndSepPunct{\mcitedefaultmidpunct}
{\mcitedefaultendpunct}{\mcitedefaultseppunct}\relax
\EndOfBibitem
\bibitem[Ruzette(2002)]{Ruzette}
Ruzette,~A.-V. \emph{Nature Materials} \textbf{2002}, \emph{1}, 85\relax
\mciteBstWouldAddEndPuncttrue
\mciteSetBstMidEndSepPunct{\mcitedefaultmidpunct}
{\mcitedefaultendpunct}{\mcitedefaultseppunct}\relax
\EndOfBibitem
\bibitem[Leeuw(1980)]{Ewald}
Leeuw,~S. W.~D. \emph{Proc.\ R. \ Soc.\ London A} \textbf{1980}, \emph{373},
  27\relax
\mciteBstWouldAddEndPuncttrue
\mciteSetBstMidEndSepPunct{\mcitedefaultmidpunct}
{\mcitedefaultendpunct}{\mcitedefaultseppunct}\relax
\EndOfBibitem
\bibitem[van Gunsteren and Berendsen(1977)van Gunsteren, and
  Berendsen]{gunsteren1}
van Gunsteren,~W.~F.; Berendsen,~H. J.~C. \emph{Mol.\ Phys.} \textbf{1977},
  \emph{34}, 1311\relax
\mciteBstWouldAddEndPuncttrue
\mciteSetBstMidEndSepPunct{\mcitedefaultmidpunct}
{\mcitedefaultendpunct}{\mcitedefaultseppunct}\relax
\EndOfBibitem
\bibitem[van Gunsteren and Berendsen(1982)van Gunsteren, and
  Berendsen]{gunsteren2}
van Gunsteren,~W.~F.; Berendsen,~H. J.~C. \emph{Mol.\ Phys.} \textbf{1982},
  \emph{45}, 637\relax
\mciteBstWouldAddEndPuncttrue
\mciteSetBstMidEndSepPunct{\mcitedefaultmidpunct}
{\mcitedefaultendpunct}{\mcitedefaultseppunct}\relax
\EndOfBibitem
\bibitem[Nan et~al.(2010)Nan, Shen, and Ma]{Nan1}
Nan,~C.-W.; Shen,~Y.; Ma,~J. \emph{Annu.~Rev.~Mater.~Res.} \textbf{2010},
  \emph{40}, 131\relax
\mciteBstWouldAddEndPuncttrue
\mciteSetBstMidEndSepPunct{\mcitedefaultmidpunct}
{\mcitedefaultendpunct}{\mcitedefaultseppunct}\relax
\EndOfBibitem
\bibitem[Goswami et~al.(2007)Goswami, Kumar, Bhattacharya, and
  Douglas]{Goswami1}
Goswami,~M.; Kumar,~S.~K.; Bhattacharya,~A.; Douglas,~J.~F.
  \emph{Macromolecules} \textbf{2007}, \emph{40}, 4113\relax
\mciteBstWouldAddEndPuncttrue
\mciteSetBstMidEndSepPunct{\mcitedefaultmidpunct}
{\mcitedefaultendpunct}{\mcitedefaultseppunct}\relax
\EndOfBibitem
\bibitem[Wang et~al.(2002)Wang, Dormidontova, and Lodge]{Lodge3}
Wang,~J.~F.; Dormidontova,~E.~E.; Lodge,~T.~P. \emph{Macromolecules}
  \textbf{2002}, \emph{35}, 9687\relax
\mciteBstWouldAddEndPuncttrue
\mciteSetBstMidEndSepPunct{\mcitedefaultmidpunct}
{\mcitedefaultendpunct}{\mcitedefaultseppunct}\relax
\EndOfBibitem
\bibitem[Ono et~al.(2007)Ono, Sugimoto, Shinkai, and Sada]{Ono1}
Ono,~T.; Sugimoto,~T.; Shinkai,~S.; Sada,~K. \emph{Nat.~Matter.} \textbf{2007},
  \emph{6}, 429\relax
\mciteBstWouldAddEndPuncttrue
\mciteSetBstMidEndSepPunct{\mcitedefaultmidpunct}
{\mcitedefaultendpunct}{\mcitedefaultseppunct}\relax
\EndOfBibitem
\bibitem[Li et~al.(2004)Li, Kesselman, Talmon, Hillmyer, and Lodge]{Lodge1}
Li,~Z.~B.; Kesselman,~E.; Talmon,~Y.; Hillmyer,~M.~A.; Lodge,~T.~P.
  \emph{Science} \textbf{2004}, \emph{306}, 98\relax
\mciteBstWouldAddEndPuncttrue
\mciteSetBstMidEndSepPunct{\mcitedefaultmidpunct}
{\mcitedefaultendpunct}{\mcitedefaultseppunct}\relax
\EndOfBibitem
\bibitem[Lodge(2008)]{Lodge2}
Lodge,~T.~P. \emph{Science} \textbf{2008}, \emph{321}, 5885\relax
\mciteBstWouldAddEndPuncttrue
\mciteSetBstMidEndSepPunct{\mcitedefaultmidpunct}
{\mcitedefaultendpunct}{\mcitedefaultseppunct}\relax
\EndOfBibitem
\end{mcitethebibliography}
\newpage
\listoffigures 
\end{document}